\documentclass[12pt]{iopart}
\usepackage{graphicx}
\usepackage{cite}
\graphicspath{{./figs-pdf/}}

\usepackage{tikz}
\usepackage{dcolumn}
\usepackage{bm}
\usepackage{graphicx, graphics, color}

\begin{document}

\title[Thermoelectric nano-devices in the Kondo regime]{Multi-terminal far-from-equilibrium thermoelectric nano-devices in the Kondo regime}

\author{Ulrich Eckern}
\address{Institute of Physics, University of Augsburg, 86135 Augsburg, Germany} 
\ead{ulrich.eckern@physik.uni-augsburg.de}
\author{Karol I.\ Wysoki\'{n}ski}
\address{Institute of Physics, M.~Curie-Sk{\l}odowska University, pl.~M.~Curie-Sk{\l}odowskiej 1, 20-031 Lublin, Poland}
\ead{karol.wysokinski@poczta.umcs.lublin.pl}


\begin{abstract}
The quest for good thermoelectric materials and/or high-efficiency thermoelectric 
devices is of primary importance from theoretical and practical points of view. 
Low-dimensional structures with quantum dots or molecules are promising candidates 
to achieve the goal. Interactions between electrons, far-from-equilibrium conditions 
and strongly non-linear transport are important factors affecting the usefulness of the devices. 
This paper analyses the thermoelectric power of a two-terminal quantum dot under large thermal $\Delta T$ 
and voltage $V$ biases as well as the performance of the three-terminal system as a heat engine. {To properly 
characterise the non-linear effects under these conditions},  
two different Seebeck coefficients are introduced, generalizing the linear response expression.
The direct calculations of thermally induced electric and heat currents show, in agreement 
with  recent work, that the efficiency of the thermoelectric heat engine as measured 
by the delivered power is maximal far from equilibrium. Moreover, the strong Coulomb interactions 
between electrons on the quantum dot are found to diminish the efficiency at maximum 
power and the maximal value of the delivered power, {both in the Kondo regime and outside of it.}
\end{abstract}

\maketitle
 
\section{Introduction}
Thermoelectricity, the invention of the 19th century, is still at the forefront 
of research due to its importance for space exploration and automotive industry 
\cite{yang2006}, and many  more branches of modern technology both at large \cite{chu2012} 
and small \cite{whitney2018} scales. Attempts to find high performance thermoelectric
bulk materials \cite{ren2018}, including those with topologically non-trivial \cite{gooth2018}
band structure, have seen limited progress. In the last few decades, a lot of attention has been 
put forward towards nano-devices \cite{benenti2017} and molecular 
systems \cite{zimbovskaya2011,russ2016}, utilizing quantum effects to boost their 
thermoelectric performance towards the thermodynamic limit \cite{josefsson2018}.

\begin{figure}
\centerline{\includegraphics[width=0.5\linewidth]{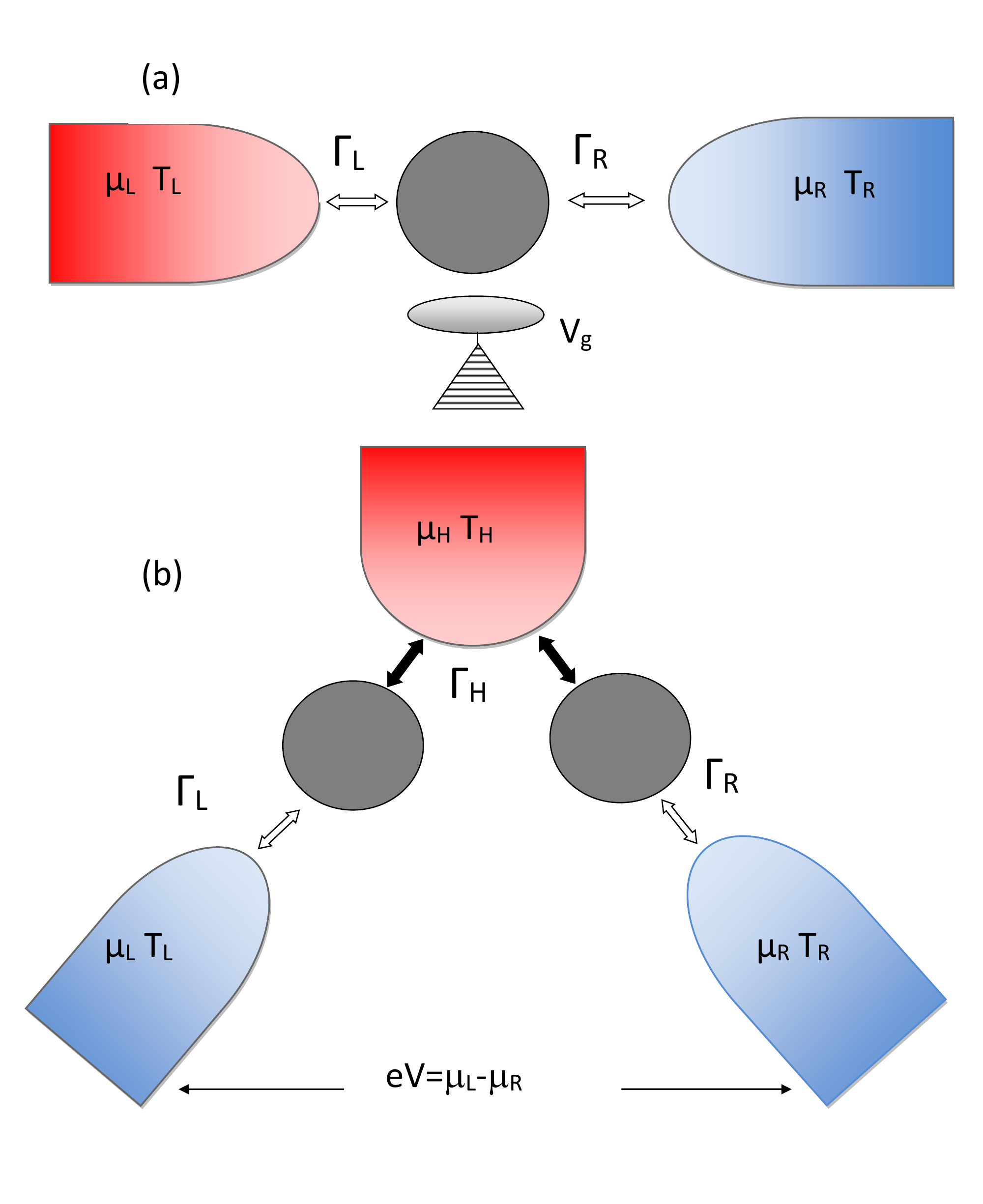}}
\caption{(color online) Two examples of devices with quantum dots. In panel (a) we show
a two-terminal quantum dot device with the left electrode at higher temperature (red). Panel (b) shows
a simple nano-engine with one hot (red) and two cold electrodes (blue) and two quantum dots (grey). 
In the latter case the filtering properties of quantum dots are important for the performance of the device.}
\label{rys1}
\end{figure}

When a temperature gradient $\nabla T$ (or a temperature difference $\Delta T$) 
is established across a bulk material, the voltage $V$ is generated. The response 
of the isotropic system is quantitatively characterized by the Seebeck
coefficient \cite{ashcroft-mermin,zlatic2014} 
\begin{equation}
S=-\left(\frac{V}{\Delta T}\right)_{I=0},
\label{seebeck}
\end{equation}
defined under the condition of zero charge current.
The same formal definition is valid for a nano-structure 
with two external electrodes (see Fig.~\ref{rys1}(a)). However, Eq.~(\ref{seebeck}) has 
to be generalized for more complicated geometries with several electrodes. 
In fact the thermoelectric characterization of nano-structures with three  
electrodes (as, {\it e.g.}, the one shown in Fig.~\ref{rys1}(b)) requires the definition of
the whole matrix of Seebeck coefficients. Such geometries also allow studies of
non-local effects \cite{mazza2014,michalek2016}.

In bulk systems the linear approximation is generally a valid simplification \cite{mahan}.
{Under such a condition the thermoelectric figure of merit, $ZT=GS^2T/\kappa$, where $G$ is 
the conductance,
$\kappa$ thermal conductance, and $S$ the Seebeck coefficient, is viewed as the most important 
factor deciding on the usefulness of the material as heat to electricity converter: namely, the 
efficiency of the thermoelectric heat to electricity converter is given 
by  $\eta=\eta_C (\sqrt{ZT+1}-1)/(\sqrt{ZT+1}+1)$}, where $\eta_C$ is the Carnot efficiency.
In nano-structures containing quantum dots, however, we are virtually always dealing with a non-linear
situation, as mentioned 
earlier \cite{muralidharan2012,whitney2013, wysokinski2016} and carefully discussed recently \cite{benenti2017}.
The small ratio of the thermalization length to the sample length in bulk systems,
and the opposite limit in nano-structures is responsible for their different behaviour. This has a
profound effect on the analysis of small heat engines, and implies that 
the thermoelectric figure of merit $ZT$ is not a useful parameter to judge the efficiency of a device.  
This also means that nano-systems with a large figure of merit \cite{finch2009} 
$ZT$ may in fact feature a small efficiency \cite{mazza2014,muralidharan2012,whitney2013,meair2013,szukiewicz2015}. 

From a basic physics point of view, the Seebeck coefficient provides 
additional and novel information about the investigated system 
compared to that obtained from the electrical conductivity. In the simplest
case, the latter depends on the value of the density of states $N(E_F)$ at the Fermi level, while
the former `measures' its slope. The thermopower has been shown to be directly related to the entropy flowing 
between different parts of the system \cite{zlatic2014}. {In fact, the entropy of 
nanosystems has been recently measured \cite{kleeorin2019} by means of thermoelectric transport.}
A strong increase of $S$ in nano-devices and nano-structured materials  
has been observed \cite{heremans2004}, in agreement with the earlier proposal \cite{mahan1996}. 
Recent studies \cite{wiendlocha2012} have shown that doping or nano-structuring bulk thermoelectric 
materials may lead to the required modifications of the energy spectrum close to the Fermi energy but
also to localization of states which deteriorates the systems performance.
The details of the nano-structures may be important, $e.g.$,
for systems based on molecules the actual value of the thermopower depends 
on the length of the molecule \cite{sadeghi2015,zimbovskaya2017}.

It has been found \cite{josefsson2018,whitney2013} that non-linear working conditions can favourably affect 
the performance of heat engines. Indeed recent years have witnessed increased activities 
in the theoretical studies of transport beyond the linear 
approximation \cite{hershfield2013,balserio2010,dutt2013,lopez2013,azema2014,muttalib2015,dorda2016}. 
The main motivation of that kind of work is related to the desire to find devices with 
improved thermoelectric performance, which often can be achieved in the non-linear regime only \cite{jordan2013}.
The early developments in non-linear quantum transport driven by thermal gradients and/or 
voltage biases have been reviewed recently \cite{sanchez2016}. Even more recent work includes 
Refs. \cite{erdman2017,karbaschi2017,jiang2017,sierra2017,karki2017,hussein2019,dutta2019}.

The aim of the present work is to study thermoelectric transport of quantum dot based
nano-devices by means of the non-equilibrium Green function approach, taking Coulomb 
interactions and non-linear effects into account.
In the non-linear regime more general definitions of the Seebeck coefficient than 
that given in Eq.~(\ref{seebeck}) are needed. These are especially important 
if an externally applied voltage $V$ is present while the system is thermally biased. 
In the presence of interaction and at low temperature, the Kondo effect is expected to dominate the
transport of the system at hand. The existence of the Kondo effect in quantum 
dots has been predicted a long time ago \cite{glazman1988,lee1988}, and later observed 
experimentally \cite{GoldhaberGordon,Cronenwett,Schmid}. 

As the theoretical treatment of an interacting quantum dot is of importance in itself, 
we also present in some detail the semi-analytical technique proposed recently by 
Lavagna \cite{lavagna2015}. It is based on the equation of motion method \cite{zubarev1960} 
for the non-equilibrium (or Keldysh) Green functions \cite{haug-jauho1996}. She has 
proposed a few important additions, which allow to properly describe the
Kondo effect \cite{hewson} in linear and non-linear regimes, {\it i.e.}, under large voltage and 
temperature differences between the electrodes and for the particle-hole symmetric model.

In the present work we extend our previous studies of non-equilibrium screening 
effects \cite{szukiewicz2016}, which is important for a better understanding of the interaction 
effects in far-from-equilibrium situations as 
encountered in {three-terminal two-quantum-dot high-efficiency heat engines \cite{jordan2013,donsa2014}.}
  
The use of the equation of motion (EOM) technique to study the single impurity Anderson 
model has a long history. It started with the work of Anderson himself \cite{anderson1961}, and 
has been pursued by others \cite{sim-appr,lacroix1981,czycholl1996,kuzemsky2002},
mainly in the context of single impurities in metals. Some of the attempts at generalizing the
original EOM and the decoupling procedures have been summarized in \cite{kashcheyevs2006}.   
The technique was later applied to the Kondo effect in quantum dots coupled to external
electrodes \cite{glazman1988,lee1988}. More recent work includes studies of more
complicated systems and geometries like those with one normal and one ferromagnetic 
or superconducting electrode 
\cite{ng1996,boese2001,dong2002,krawiec2006,galperin2007, domanski2008,monreal2009,swirkowicz2009,costi2010,vanroermund2010,michalek2013,smirnov2013,zimbovskaya2014,zimbovskaya2015,gorski2015,silva2017,daroca2018}.

It has to be mentioned that there is a vast amount of experimental data measuring the thermopower in various nano-systems 
\cite{molenkamp1992,staring1993,dzurak1993,dzurak1998,godijn1999,harman2002,tian2012,yanagi2014,wu2013,llaguno2003,scheibner2005,pogosov2006,reddy2007,scheibner2007,chang2006,scheibner2008,saira2007,svensson2012,svensson2013,thierschmann2013,matthews2014,roche2015,hartmann2015,thierchmann2015,svilans2016,svilans2016a,svilans2018}; 
these provide additional motivation for the theoretical work.  
{Very recently the three terminal two quantum dot system identical to one of the systems studied here has been
found \cite{jaliel2019} to generate a power of $0.13$ fW with efficiency of $\approx 0.15$ of the Carnot value.}

The organization of the paper is as follows. In Sec.~\ref{sec:curr} we present the model and our
approach of calculating the charge and heat currents by means of the Keldysh non-equilibrium 
Green function (GF) technique. The resulting spectral function of the quantum dot is discussed 
in Sec.~\ref{sec:kondo} at low temperatures and for various conditions including particle-hole
symmetry and non-equilibrium. The thermally induced currents are the subject of Sec.~\ref{rect}
where also rectification properties of the non-symmetrical device are mentioned. Three possible 
definitions of the Seebeck coefficients, including two valid in the non-linear regime, are 
proposed in Sec.~\ref{sec:nltep}. The  Coulomb interaction between electrons 
on the three-terminal quantum dot heat engine is found (Sec.~\ref{sec:nlheat}) 
to diminish the performance of the device in question. The appearance of the Kondo effect 
in the heat engine shows up as a two-leaf structure of the performance diagram on the efficiency-power plane. 
We end with  summary and conclusions in Sec.~\ref{sec:sum}.

\section{Modeling the device and calculating currents}\label{sec:curr}
Here we discuss the simple geometry where the system consists of a quantum dot tunnel-coupled to
a few normal electrodes as shown in Fig.~\ref{rys1}. The Hamiltonian of the system is written as
\begin{equation}
{H}=\sum_{\lambda,{k},\sigma}\varepsilon_{\lambda {k}}n_{\lambda {k} \sigma} + 
\sum _{\sigma} \varepsilon _{\sigma}n_{\sigma} +Un_\uparrow n_\downarrow 
+\sum_{\lambda {k} \sigma} ({V}_{\lambda {k}\sigma} c^{\dagger}_{\lambda k\sigma} d_{\sigma} 
+ {V}_{\lambda {k}\sigma}^* d^{\dagger}_{\sigma} c_{\lambda {k} \sigma}),
\label{ham1}
\end{equation}
where $n_{\lambda {k} \sigma}=c^{\dagger}_{\lambda {k} \sigma}c_{\lambda {k} \sigma}$
and $n_{\sigma}=d^{\dagger}_{\sigma} d_{\sigma}$
denote particle number operators for the leads and the dot, respectively.  
The operators $c^{\dagger}_{\lambda k\sigma} (d^{\dagger}_{\sigma})$ create electrons in
respective states $\lambda {k}\sigma$ $(\sigma)$ in the leads $\lambda = L, R, H$ (on the dot). 
{The wave vector $k$ denotes the Bloch state in the electrodes}, the spin is $\sigma=\pm 1 \; (\uparrow, \downarrow)$, 
and $\varepsilon_\sigma=\varepsilon_d+\sigma \mu_B B$, 
where $B$ is the magnetic field introducing Zeeman splitting, $\mu_B$ denotes the Bohr magneton, 
and $\varepsilon_d$ is the dot electron energy level.
The Hubbard parameter $U$ describes the repulsion between two electrons on the dot. 

The charge current in the electrode $\lambda$ is calculated
as the time derivative of the average charge in that electrode
$\langle N_{\lambda}\rangle=\sum_{k\sigma} \langle n_{\lambda{k}\sigma}\rangle$, and reads 
\begin{equation}
I_{\lambda}=-e\left\langle\frac{dN_{\lambda}}{dt}\right\rangle=\frac{ie}{\hbar}\langle [N_\lambda,\hat{H}]\rangle
\label{current1}
\end{equation}
where $\langle...\rangle$ denotes the statistical average.
The calculation of the heat fluxes, $J_\lambda$, follows that of the charge:
\begin{equation}
J_\lambda=\frac{i}{\hbar}\langle [H_\lambda,\hat{H}]\rangle -\mu_\lambda\frac{i}{\hbar}\langle [N_\lambda,\hat{H}]\rangle,
\label{hcurrent}
\end{equation} 
where
$H_\lambda=\sum_{{k},\sigma}\varepsilon_{\lambda {k} \sigma}n_{\lambda {k} \sigma}$ 
is the energy operator for the electrode $\lambda$. 
Calculating the commutators and introducing appropriate GFs, one obtains\cite{haug-jauho1996}  
\begin{eqnarray}
I_{\lambda}(t)=\frac{2e}{\hbar}\sum_{{k}\sigma}{\rm Re}
\bigg[{V}_{\lambda{k}\sigma}G^{<}_{\sigma,\lambda{k}\sigma}(t,t)\bigg],
\label{curr1} \\
J_{\lambda}(t)=\frac{2e}{\hbar}\sum_{{k}\sigma}(\varepsilon_{\lambda {k}}-\mu_\lambda) {\rm Re}
\bigg[{V}_{\lambda{k}\sigma}G^{<}_{\sigma,\lambda{k}\sigma}(t,t)\bigg].
\label{hcurr1}
\end{eqnarray}
The final expressions for the (stationary) currents can easily be written in the following general 
form: \cite{haug-jauho1996}
\begin{eqnarray}
I_{\lambda}&=&\frac{ie}{\hbar} \int\frac{dE}{2\pi}\sum_\sigma \Gamma_\sigma^{\lambda}(E)
\{ G_\sigma^{<}(E) + f_{\lambda}(E) [G_\sigma^{r}(E)-G_\sigma^{a}(E)]\}, 
\label{charge-curr} \\
J_{\lambda}&=&\frac{ie}{\hbar} \int\frac{dE}{2\pi}\sum_\sigma\Gamma_\sigma^{\lambda}(E)
(E-\mu_\lambda)\{ G_\sigma^{<}(E) + f_{\lambda}(E) [G_\sigma^{r}(E)-G_\sigma^{a}(E)]\}.
\label{heat-curr}
\end{eqnarray}
The parameters $\Gamma_\sigma^{\lambda}(E)=2\pi\sum_{{k}}|V_{\lambda{k}\sigma}|^2\delta(E-\varepsilon_{\lambda{k}})$
describe the coupling between the dot and the respective electrodes.
The Green functions $G^{i}_\sigma(E)=\langle  \langle  d_\sigma|d^\dagger_\sigma\rangle\rangle^i_{E}$ 
with $i=r,a,<$ determine the spectral properties of the quantum dot as well as the transport properties 
of the whole system. 

Having in mind non-equilibrium transport induced by a voltage bias and/or a temperature difference,
we keep the dependence of the Fermi distribution functions $f_\lambda(E)$
on the electrode $\lambda$ {\it via} its chemical potential $\mu_\lambda$ and its temperature $T_\lambda$. 
The heat current (\ref{heat-curr}) can be written as the difference between
the energy current $J^E_\lambda$ and the charge current $I_\lambda$:
\begin{equation}
J_\lambda=J^E_\lambda-\mu_\lambda I_\lambda.
\end{equation}
Importantly, in the steady state and in the wide band approximation ($i.e.$, for energy 
independent couplings: $\Gamma_\sigma^{L,R}(E)=\Gamma_\sigma^{L,R}$),
one can derive \cite{lavagna2015} {\it exact} expressions for the currents. 
First, from $\langle n_\sigma \rangle=\langle c^\dagger_\sigma(t)c_\sigma(t)\rangle$ 
and the definition of the lesser Green function, one has \cite{haug-jauho1996} 
\begin{equation}
\langle n_\sigma \rangle=-i\int \frac{dE}{2\pi}G^{<}_\sigma(E).
\label{n-dot}
\end{equation}
The derivation then makes use of the fact that in the steady state 
\begin{equation}
0=\frac{d\langle n_\sigma \rangle}{d t}=\langle \frac{d n_\sigma}{d t}\rangle=-\langle \frac{i}{\hbar} [n_\sigma,H]\rangle.
\label{n-av} 
\end{equation}
Working out the commutator in (\ref{n-av}), using (\ref{n-dot}) and the Langreth theorem \cite{langreth1976},
it is straightforward to derive the following `self-consistency' condition \cite{lavagna2015}:
 \begin{equation}
 \langle n_\sigma \rangle= i\int \frac{dE}{2\pi}
\frac{\Gamma_\sigma^L f_L(E)+\Gamma_\sigma^R f_R(E)}{\Gamma_\sigma^L+\Gamma_\sigma^R}
[G_\sigma^r(E)-G_\sigma^a(E)]. 
\label{n-noneq} 
\end{equation} 
Let us underline again that the expression (\ref{n-noneq}) is exact under the proviso that
the couplings to the leads are energy independent, $\Gamma_\sigma^{\lambda}(E) \equiv \Gamma^{\lambda}_\sigma = \mathrm{const}$.
If this condition is violated, as it might be the case in graphene \cite{wysokinskimm2012,wysokinskimm2013,wysokinskimm2014}, 
hybrid systems with one (or both) of the electrodes being a superconductor, 
{\it e.g.}, d-wave \cite{polkovnikov2002}, other approaches are needed.
For models with energy dependent couplings one still has to rely on approximate 
relations between the lesser self-energy $\Sigma^{<}_\sigma (E)$ and the retarded 
one, $\Sigma^{r}_\sigma (E)$, making use of Ng's approximation \cite{ng1996}, 
a generalisation of the fluctuation-dissipation theorem \cite{balserio2010,sierra2017}, or
using the equation of motion for the lesser GF \cite{niu1999} and suitable decoupling.
For recent attempts including energy-dependent couplings, see the paper \cite{silva2017}. 

With the above (exact in the steady state and for constant $\Gamma$'s) formulation, the charge current 
across the two-terminal system can be written in terms of the retarded GF only:
\begin{equation}
I =\frac{2e}{\hbar} \sum_\sigma \tilde{\Gamma}_\sigma \int\frac{dE}{2\pi} 
[f_L(E)-f_R(E)] \mathrm{Im} G^r_\sigma(E) ,
\label{curr-wbl}
\end{equation}
where $\tilde{\Gamma}_\sigma = \Gamma_\sigma^L\Gamma_\sigma^R / (\Gamma_\sigma^L+ \Gamma_\sigma^R)$. 
{We stress that the above formula is an exact representation of the current in terms of 
 the retarded GF of the interacting Hamiltonian, which have to be calculated either numerically 
or analytically. Here we adopt the latter approach, though obviously 
calculating the GF analytically requires some approximations, see the details given in the Appendix.
In brief, the decouplings we are using ensure that the
GFs are formally exact up to second order in the couplings. Additionally, we correct the
result by phenomenologically introducing the lifetimes \cite{lavagna2015} of various states, which
is important to capture the Kondo effect even in the particle-hole symmetric model.}

The expression (\ref{curr-wbl}) is analogous to the well-known Meir-Wingreen formula \cite{meir1992}.
A direct calculation leads to $I=I_L=-I_R$, which expresses the current conservation.
Similar expressions can be derived for the heat current flowing from the left,  
\begin{equation}
J_L=\frac{2e}{\hbar} \sum_\sigma \tilde{\Gamma}_\sigma \int\frac{dE}{2\pi} (E-\mu_L)[f_L(E)-f_R(E)] \mathrm{Im} G^r_\sigma(E),
\label{h-curr-wbl}
\end{equation}
and right electrode:
\begin{equation}
J_R=\frac{2e}{\hbar} \sum_\sigma \tilde{\Gamma}_\sigma \int\frac{dE}{2\pi} (E-\mu_R)[f_R(E)-f_L(E)] \mathrm{Im} G^r_\sigma(E).
\end{equation}
It can be verified that
\begin{equation}
J_L+J_R+(\mu_R-\mu_L)I=0
\label{en-cons}
\end{equation}
in agreement with energy conservation. Here $I=I_L=-I_R$ is the charge current, and 
$\dot{Q}=J_L+J_R$ the total heat current leaving the leads. For later use we define 
the power, $P=(\mu_R-\mu_L)I/e$, and the voltage bias, $V = (\mu_R-\mu_L)/e$, across the system. 

Three-terminal quantum dot devices have been proposed \cite{jordan2013,hershfield2013,mazza2014,szukiewicz2015}
as efficient, easy to control heat engines, and analysed in equilibrium and beyond, both for non-interacting 
and interacting quantum dots. Our general formulas for the charge and energy currents (\ref{heat-curr}) flowing out of an 
arbitrary electrode $\lambda$ allow for an easy application to the three-terminal system like the one shown 
in Fig.~\ref{rys1}(b).
In the notation of the figure the heat is flowing from the hot to two cold electrodes; assuming that
no charge current flows out of or into the hot electrode. The flow of charge is dictated by the gate bias of 
the two quantum dots, which in the discussed system act as energy filters. The total heat current in the system is 
\begin{equation}
J=J_H+J_L+J_R,
\end{equation}
while the charge current (assuming $I_H=0$) reads
\begin{equation}
I=I_L+I_R.
\end{equation}
Introducing a `load' between the two cold electrodes, in the figure shown as an external voltage $V$, one
defines the power delivered by the system as
\begin{equation}
P=IV,
\end{equation}
and the system energy harvesting efficiency as
\begin{equation}
\eta=\frac{IV}{J_H}.
\end{equation}

To calculate currents we only require the retarded Green function, which is still a complicated issue
with no exact analytic solution available. There exist a few numerically exact approaches: for example,
among the many techniques applied to study the single-impurity Anderson model \cite{hewson} the
quantum Monte Carlo \cite{hirsch1986} and the numerical renormalization group method \cite{bulla2008} 
should be mentioned. However, our aim is to use the analytic expressions which have been proven 
to be quantitatively correct \cite{lavagna2015}, and valid far from equilibrium; in addition, in our opinion
these are physically more transparent compared to purely numerical results, as discussed below.

One finds the following final expression for the on-dot GF:
\begin{equation}
\label{sol-gf}
\langle\langle d_\sigma|d^\dagger_\sigma\rangle\rangle_E= 
\frac{1+I_d(E)[\langle n_{\bar{\sigma}}\rangle+b_{1\bar{\sigma}}-b_{2\bar{\sigma}} ]}
{E-\varepsilon_\sigma-\Sigma_{0\sigma}+I_d(E)[\Sigma_1^T+\Sigma_2^T-(b_{1\bar{\sigma}}-b_{2\bar{\sigma}})\Sigma_{0\sigma} ]}, 
\end{equation}
where
\begin{equation}
I_d(E)=\frac{U}{E-\varepsilon_\sigma-U-\Sigma_{0\sigma}-\Sigma_\sigma^{(1)}-\Sigma_\sigma^{(2)}}.
\label{sol-sp2}
\end{equation}
Details and definitions are given in the Appendix.
The expression for the Green function (\ref{sol-gf}) agrees with that obtained earlier \cite{lavagna2015}. 

It has been proposed by Lavagna \cite{lavagna2015} that for the correct description of the Kondo effect
one has to supplement the above set of equations by two ingredients, see Appendix. 
First, one should introduce
the finite lifetimes of singly and doubly occupied states on the dot. Second, it is important to 
take care of the many-body renormalization of the dot energy level $\varepsilon_d$.
The first improvement consists in replacing the $E+i0$ terms in the definitions
of various Green and correlation functions by $E+i\tilde{\gamma}_\alpha$, where the inverse lifetimes $\tilde{\gamma}_\alpha$
of the excited states $\alpha=|\sigma\rangle,|\uparrow,\downarrow\rangle$ are due to higher 
order processes, and can be calculated up to the desired order {\it via} the
generalized Fermi rule as
\begin{equation}
\tilde{\gamma}_\alpha=2\pi\sum_{|f\rangle}|\langle T(E_\alpha)\rangle |^2\delta(E_\alpha-E_f),
\label{gen-fermi-rule}
\end{equation}
with $T(E)=\hat{V}+\hat{V} g(E) \hat{V} +\cdots$ being the scattering matrix, and $\hat{V}$ denoting the
part of the Hamiltonian describing the coupling between quantum dot and reservoirs.
The second improvement amounts to the replacement of $\varepsilon_d$ by $\tilde{\varepsilon}_d$, to be calculated
self-consistently from 
\begin{equation}
\tilde{\varepsilon}_d=\varepsilon_d+\Sigma_1^T(\tilde{\varepsilon}_d)+\Sigma_2^T(\tilde{\varepsilon}_d).
\label{edn}
\end{equation}
The GF (\ref{sol-gf}) has been shown to fulfil the unitarity limit and describe 
quantitatively correctly the Kondo effect \cite{lavagna2015} even of particle-hole symmetric systems
in equilibrium and out-of-equilibrium.

\section{Kondo effect in equilibrium and far from equilibrium}\label{sec:kondo}
\begin{figure}
\begin{center}
\includegraphics[width=0.47\linewidth]{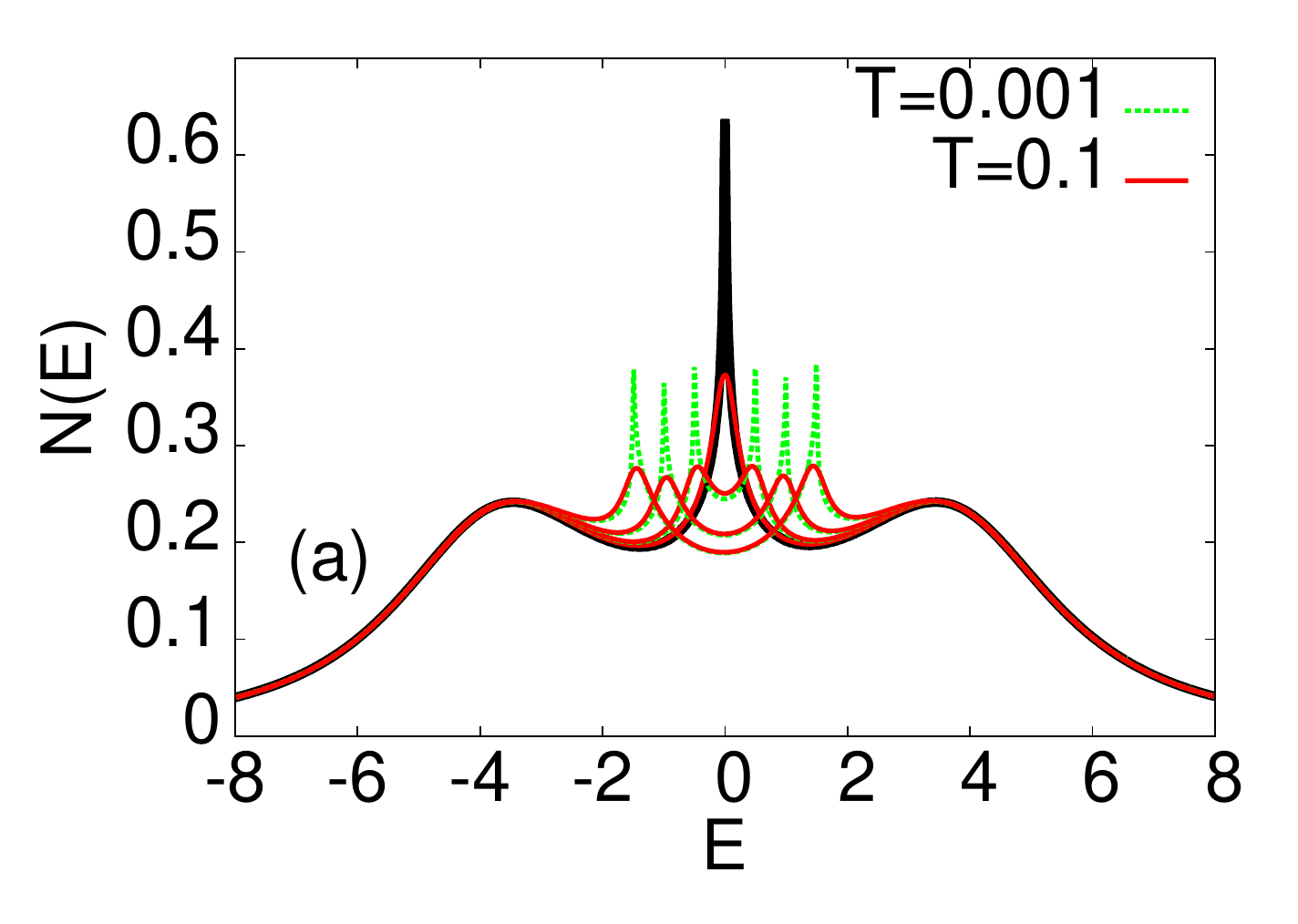}
\includegraphics[width=0.47\linewidth]{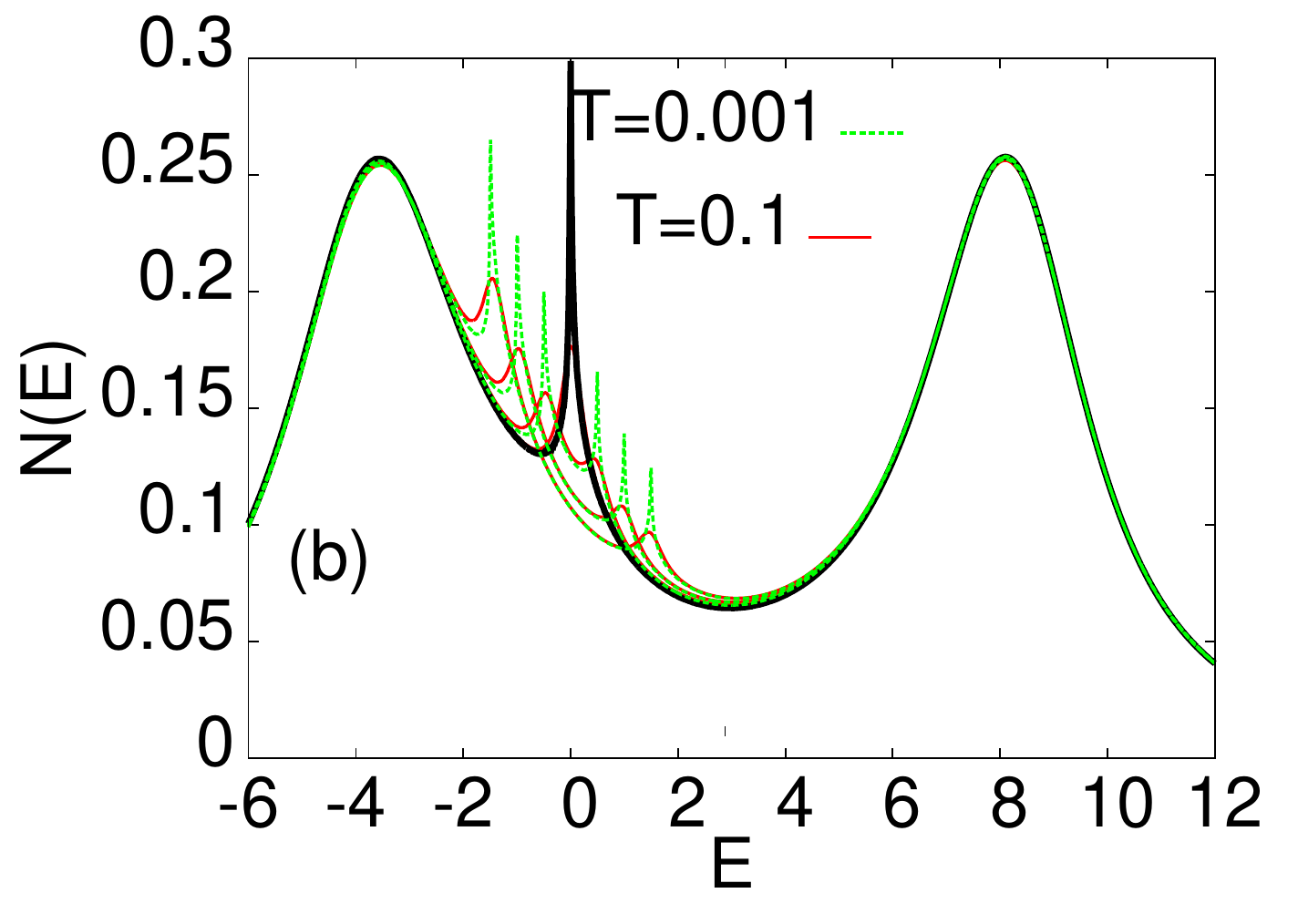}
\end{center}
\caption{(color online) The density of states (DOS) for the particle-hole symmetric interacting quantum dot 
with $\varepsilon_d=-4$, $U=8$, and $T=0.1>T_{K0}=0.061$ (a),  
and for $\varepsilon_d=-4$, $U=12$ ($T_{K0}\approx 0.0263$) (b). In both panels the DOS is given for
two values of the system temperature, $T=0.1$ and $0.001$, and for a number  
of voltages $V$ applied between left and right electrode. One observes the evolution of the Kondo peak with
increasing bias. The black (solid) curve in both parts of the figure refers to $V=0$ and $T=0.001$, and the other curves 
to increasing values of $V$. Energies are given in units of $\Gamma_0$. Note that in most formulas we set $k_B =1$, 
except when it serves clarity.}
\label{rys2}
\end{figure}

For illustrative purposes and in order to introduce the framework, we discuss in this section the properties
of an interacting quantum dot between two normal electrodes, as illustrated in Fig.~\ref{rys1}(a).
We start the presentation of the results by showing the density of states (DOS) of the quantum dot
in various regimes.
From now on, we neglect the spin dependence of the couplings, and slightly change the notation:
$\Gamma^L_{\uparrow} = \Gamma^L_{\downarrow}=\Gamma_L$, $\Gamma^R_{\uparrow} = \Gamma^R_{\downarrow}=\Gamma_R$.
In addition, we measure all energies in units of $\Gamma_0 \equiv \Gamma_L$.
The particle-hole symmetric case is of special importance as it is
well known \cite{lavagna2015} that all previous attempts to use the
EOM method failed in this case \cite{sierra2017,domanski2008,michalek2013}. In Fig.~\ref{rys2} we present the
DOS in the particle-hole symmetric situation with $\varepsilon_d=-4\Gamma_0$
and $U=8\Gamma_0$. Both lower and upper Hubbard bands centered in energy around $\varepsilon_d$ and $\varepsilon_d+U$
are clearly visible. At the same time, the Abrikosov-Suhl, sometimes also called Kondo resonance 
develops at the chemical potential. 

\begin{figure}
\begin{center}
\includegraphics[width=0.47\linewidth]{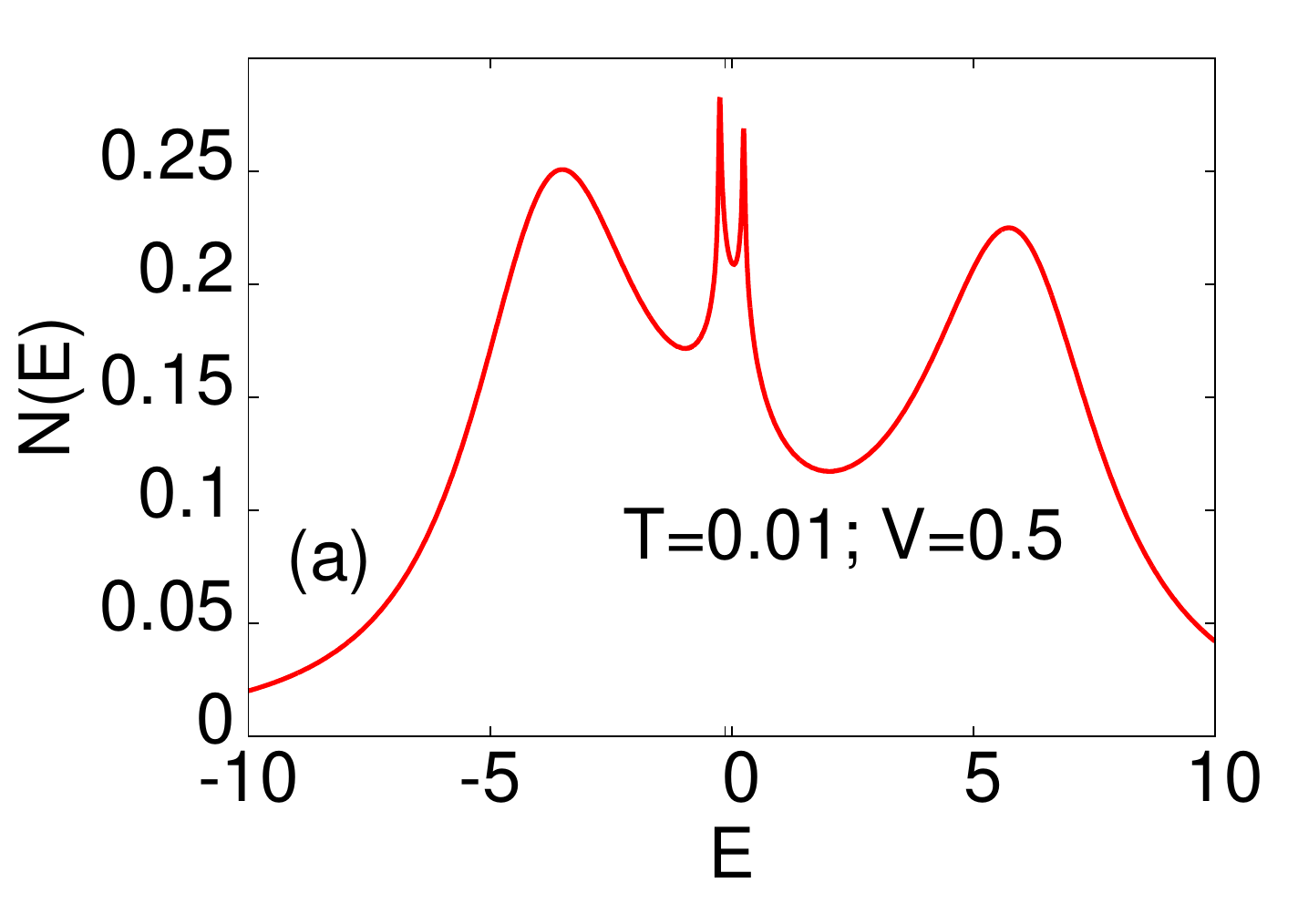}
\includegraphics[width=0.47\linewidth]{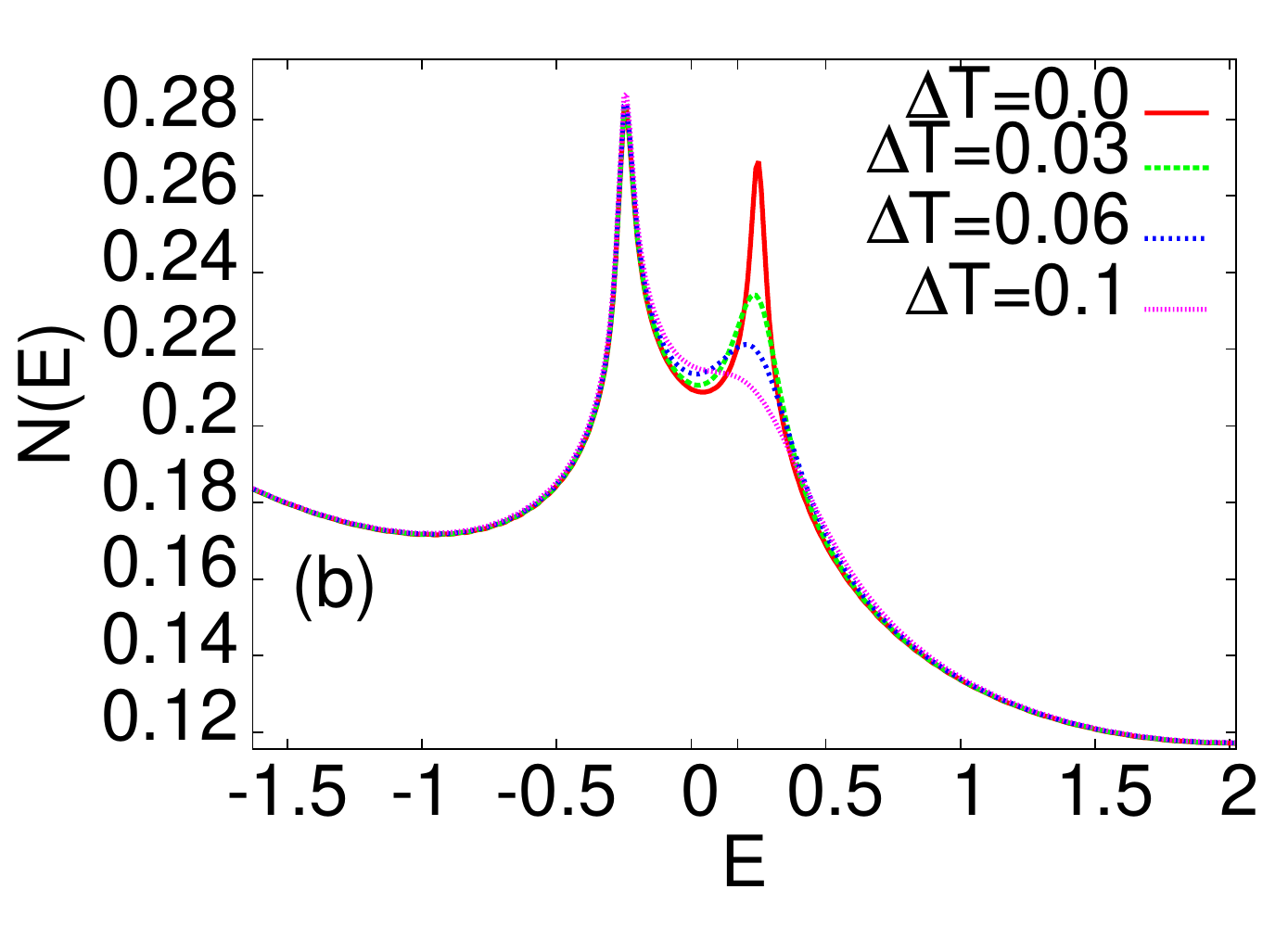}
\end{center}
\caption{(color online) DOS of the quantum dot for the voltage 
$V=0.5$ applied between left and right electrode (a), showing two Kondo peaks pinned at the
chemical potentials of the leads $\mu_{L/R}=\pm eV/2$. In panel (b) the evolution of the Kondo 
peaks with increase of temperature in the left lead by $\Delta T$ is illustrated. 
The other parameters are:  $\varepsilon_d=-4$, $U=10$, $T=0.01$, 
and $\Gamma_R=1$, again all in units of $\Gamma_0 \equiv \Gamma_L$.}
\label{rys3}
\end{figure}

The external bias shifts the chemical potentials of the leads $\mu_{L/R}=\mu\pm eV/2$ 
to new positions, and the Kondo peak splits into two with each resonance pinned to the chemical 
potential of the lead. In particular, Fig.~\ref{rys2} shows the evolution of the two peaks 
with temperature. At $T=0.001 \Gamma_0$, they are very sharp, 
while at $T=0.1\Gamma_0$ they are reduced but still clearly visible. The changes of the Kondo 
resonance with bias and temperature are crucial to understand the behaviour of the thermally induced
current and the (non-linear) thermoelectric power as discussed in the next sections.
At still higher temperatures (not shown) both Kondo peaks vanish altogether, and only lower and upper Hubbard
bands survive. The Kondo resonance is due to spin flip processes on the dot while the Hubbard sub-bands are related
to charge fluctuations. This explains the relative robustness of Hubbard sub-bands with increasing temperature,
and the fragility of the Kondo correlations. However, voltage and temperature affect 
the Kondo resonance in a different way. While a voltage $eV>T_K$ leads to a splitting of the resonance in two
sub-peaks with concomitant decrease of their maximum, the finite temperature only lowers the height of the peak 
and broadens it. All these features are  well reproduced by the used decoupling scheme.  
Outside the particle-hole symmetric point, the voltage-split Kondo resonances are not
symmetric anymore. This is mainly due to the closer proximity of one of the resonances to the lower 
Hubbard band. The individual resonances are pinned to the Fermi levels of the respective electrodes. This is 
well visible in  Fig.~\ref{rys2}(b), and also in  Fig.~\ref{rys3}(a).

We now add a thermal bias to one of the electrodes, with focus on the question of how 
the density of states evolves with temperature difference. 
We start with the split Kondo resonance as shown in Fig.~\ref{rys3}(a). 
The temperature of the right lead is kept constant, at 
$T_R=T$, while we gradually increase the temperature of the left electrode, $T_L=T+\Delta T$.
The  panel (a) of the Fig.~\ref{rys3} shows the voltage-split Kondo peaks for $\Delta T =0$, for
easy comparison, while the  panel (b) demonstrates the evolution of both peaks with increasing $T_L$.
The peak pinned to the chemical potential of the left electrode (appearing at $E=\mu_L$) is strongly 
affected by the temperature bias, it broadens and vanishes with increasing temperature. One 
observes only small changes of the other Kondo peak. Thus under voltage and temperature bias 
the peaks' heights and widths depend mainly on the 
conditions ($V$, $\Delta T$) at the electrode  with the chemical potential 
to which it is pinned. For diminishing external voltage bias both  
peaks overlap in energy, and one observes a single feature in the density of states
{corresponding} to the average temperature $(T_L+T_R)/2$ of the system.

\section{Thermocurrents and their rectification} \label{rect}
{Recently there has been great interest in the control of heat 
transport \cite{li2012} and its use to process information in analogy to electronics.  
To this end, one typically uses physical systems which are driven by a temperature bias. The field 
develops fast with many theoretical proposals and related experimental realisations 
\cite{chang2006,scheibner2008,ruokola2009,kuo2010,diaz2011,rurali2014,craven2018,suarez2018}. For 
a recent overview of the experimental advances
in thermal rectification with particular attention to nanoscale devices, the reader 
is invited to consult the recent review \cite{liu2019}, and references therein.}

\begin{figure}
\begin{center}
\includegraphics[width=0.530\linewidth]{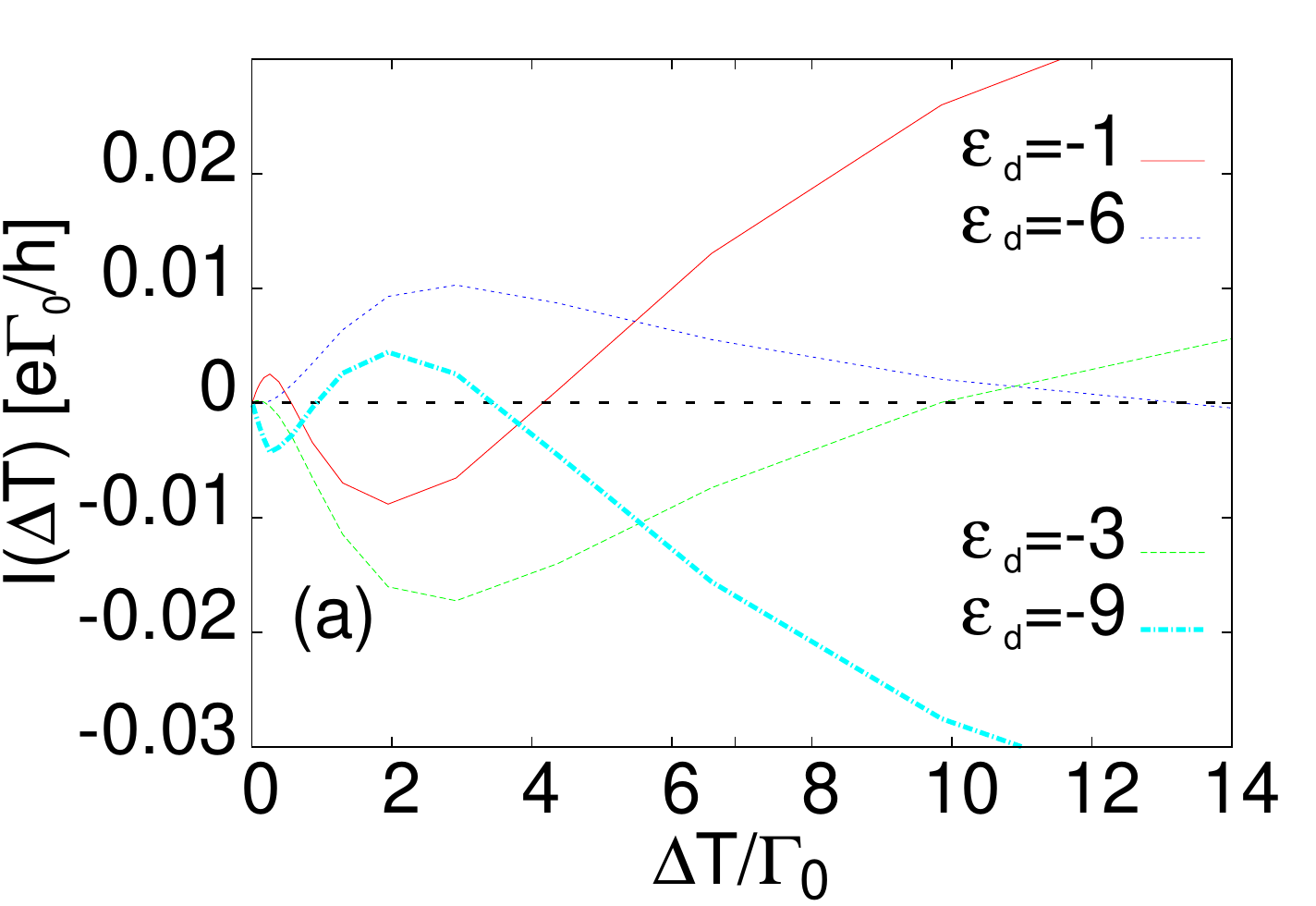}
\includegraphics[width=0.390\linewidth]{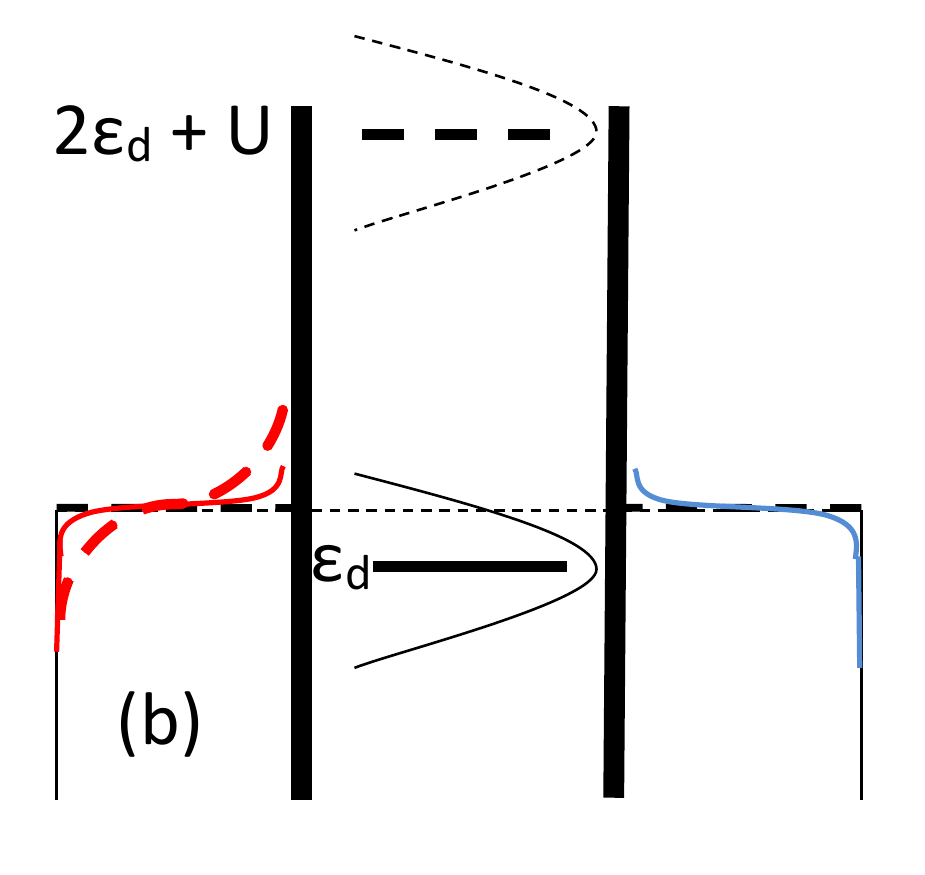}
\end{center}
\caption{(color online) The thermally induced  current ($V=0$) {\it vs.} the temperature difference $\Delta T$ 
in a system with $T_R=T=0.01\Gamma_0$, and $T_L=T+\Delta T$, for a few values of $\varepsilon_d$ and $U=10\Gamma_0$
is shown in panel (a). Panel (b) gives a schematic (not to scale) energy diagram for
$\varepsilon_d=-\Gamma_0$. The blue curve on the right illustrates the Fermi distribution function corresponding 
to $T_R=T$, while the red (thin) line corresponds to $T_L=T+\Delta T$ for infinitesimally small $\Delta T$; 
the red dashed line corresponds to $\Delta T\approx \Gamma_0$. In the latter situation, the holes 
flow from left to right (or electrons in the opposite direction), and this leads to a negative thermo-current 
as shown in the left panel. The dot density of states around $\varepsilon_d$ and $\varepsilon_d+U$ are 
shown very schematically (without the Kondo resonance appearing at very low temperatures at energy $E = \mu$).}
\label{rys4}
\end{figure}

{Motivated by the importance of thermal rectification and the fact that it is typically induced by
the temperature difference across the nanodevice, we first} 
calculate the thermo-currents, {\it i.e.}, the thermally induced currents, and {later focus on
rectification phenomena in systems with broken symmetry. In Fig.~\ref{rys4} we show the current}  
$I(T,\Delta T) = I(T,\Delta T, V=0)$ for the interacting system with $U=10\Gamma_0$,  
$T=0.01\Gamma_0$, and  for a few values of $\varepsilon_d$. 
As a function of temperature bias $\Delta T$ the thermo-current shows an interesting 
set of zero values. The number of zeros depends on 
the interaction strength and the gate voltage, {\it i.e.}, $\varepsilon_d$. To understand 
this behaviour let us consider a particular value of $\varepsilon_d$. For 
$\varepsilon_d=-\Gamma_0$ and the assumed $U$, the Kondo temperature is $T_K\approx 0.09\Gamma_0$. 
In such a situation the on-dot level is slightly below the chemical potential 
$\mu=\mu_L=\mu_R$, and the doubly occupied state $2\varepsilon_d+U=8\Gamma_0$ is far above it, 
so the Kondo resonance develops. The schematic energy diagram (without Kondo resonance) is
illustrated in  Fig.~\ref{rys4}(b). For a small average temperature, the temperature 
difference $\Delta T$ induces the current flowing mainly {\it via} the Kondo resonance 
(located at $E\approx \mu=0$) out of the left electrode, and $I(\Delta T) >0$. 
For a temperature difference around $\Delta T \approx \Gamma_0 >T_K$, the contribution 
from the smeared Kondo resonance gets relatively smaller than the contribution from the other states. 
For the considered, relatively large temperature difference  
it is mainly holes which flow from the left electrode or 
electrons mainly flowing from the right to the left due to the difference of the Fermi functions.
For still higher $\Delta T$, the thermally excited electrons at the left electrode start 
to flow from left to right through the upper doubly occupied level, and the current is again positive.
In a similar manner one can understand the behaviour of the current 
at other gate voltages \cite{sierra2017}, and other $\varepsilon_d$. {The effect has been
thoroughly discussed recently \cite{sanchez2016}, and measured experimentally \cite{svensson2013}}.

According to Eqs.~(\ref{curr-wbl}) and (\ref{h-curr-wbl}), the charge and heat  
currents through the quantum dot directly depend on the on-dot Green function.
In the linear approximation the matrix of kinetic coefficients is symmetric, according to Onsager's relations.
Outside linearity one expects the violation of these symmetries, and a non-linear
dependence of the currents on the voltage and/or thermal bias. {An even more interesting behaviour is expected 
if the device shows an internal asymmetry.} The asymmetry of a device induces current
rectification effects as discussed earlier in the context of transport {\it via} 
molecular junctions in the pair-tunnelling regime \cite{wysokinski2010}. In the two-terminal 
structure with the same spectrum of the left and right electrode the only source of asymmetry 
are the couplings to the leads $\Gamma_{L(R)}$. 
\begin{figure}
\includegraphics[width=0.470\linewidth]{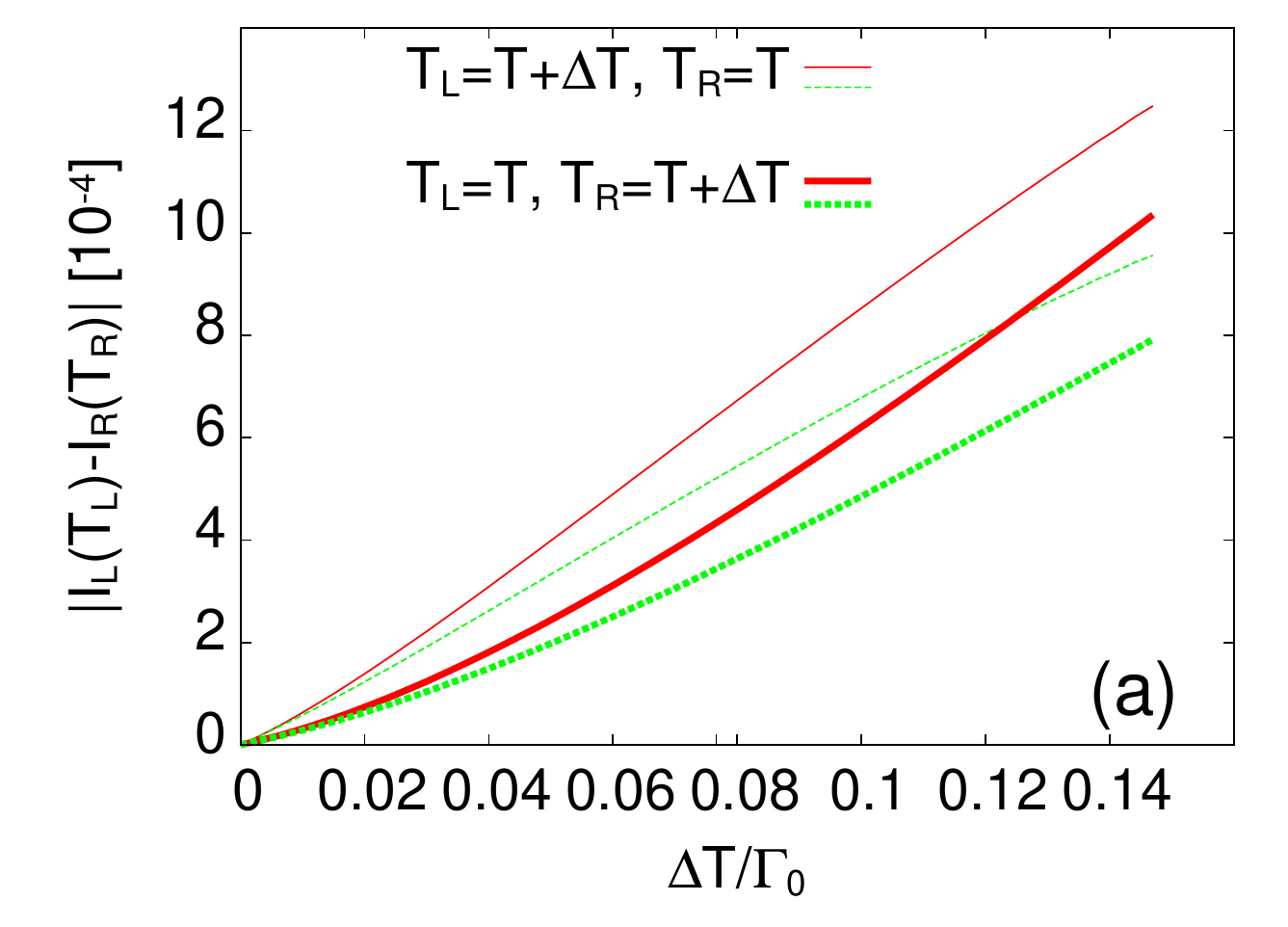}
\includegraphics[width=0.470\linewidth]{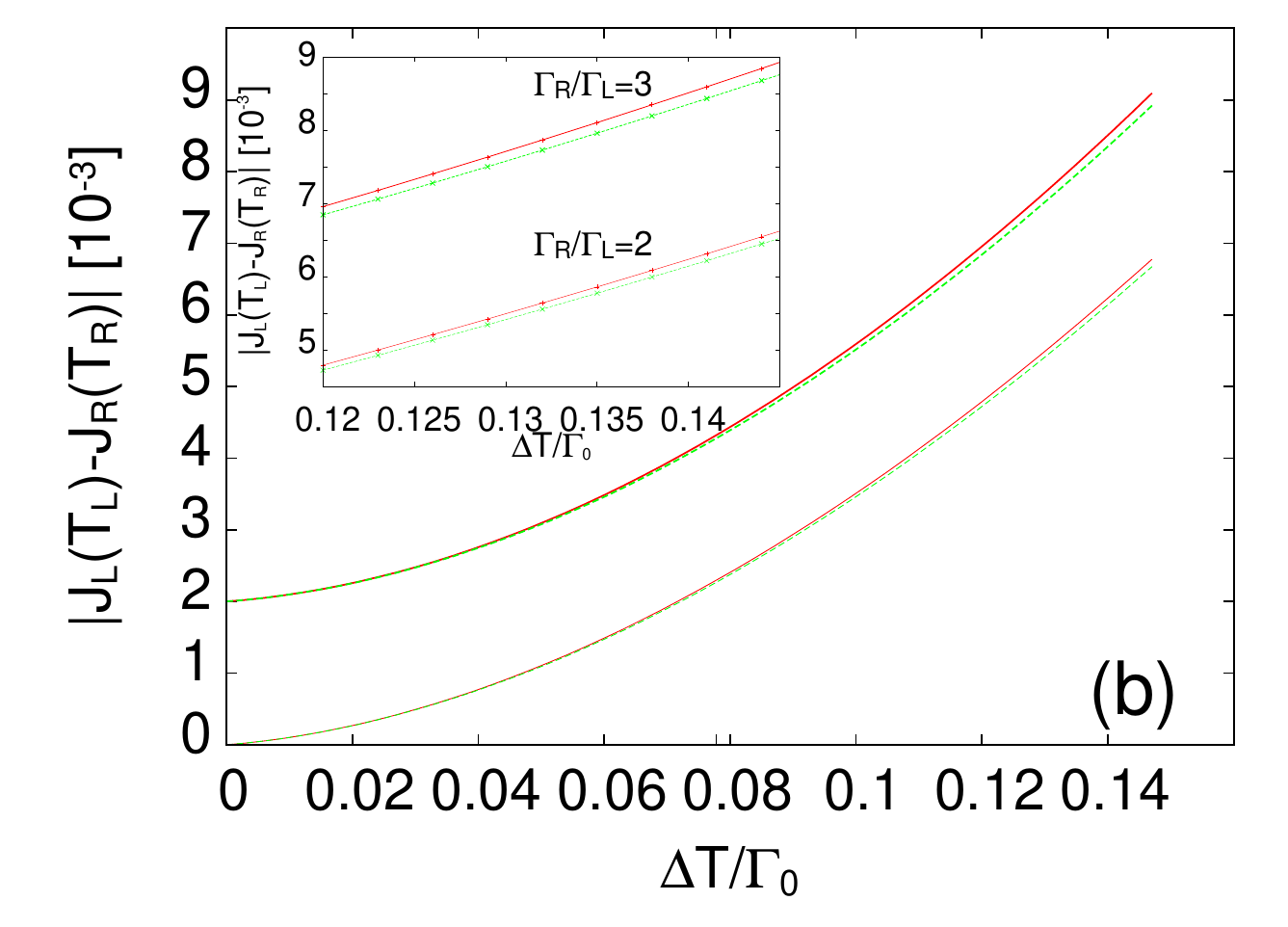}
\caption{(color online) (a) Thermally induced charge currents, $I(T,\Delta T)=|I_L(T_L)-I_R(T_R)|$, 
in systems with asymmetric couplings calculated for
$\Delta T$ applied to the left electrode (red curves), and the same $\Delta T$ applied
to the right electrode (green curves) $vs.$ temperature difference $\Delta T$, 
with $T=0.01\Gamma_0$, $V=0$, $\varepsilon_d=-2\Gamma_0$,
and $U=8\Gamma_0$. Thin (thick) curves correspond to the asymmetry $\Gamma_R /\Gamma_L=2$
($\Gamma_R /\Gamma_L=3$). (b) Thermally induced heat currents, $|J_L(T_L)-J_R(T_R)|$,
and their rectification. The colour code is the same as in panel (a), but now the curves for
$\Gamma_R/\Gamma_L=3$ are shifted vertically by $+2\cdot 10^{-3}$ for better visibility. The thermal
rectification is an order of magnitude smaller. The inset shows the $\Delta T >0.12\Gamma_0$ part of
the main figure. }
\label{rys5}
\end{figure}
To show the thermally induced charge and heat current rectification in the Kondo regime, we assume an asymmetry in the 
couplings with $\Gamma_R/\Gamma_L=2,3$. { We apply first the temperature bias $\Delta T$ to the left electrode and calculate 
the current across the system, $I(T,\Delta T)=I_L(T_L)-I_R(T_R)$, and later apply the same temperature difference to
the right electrode, keeping the average temperature of the system constant}. For the studied temperature range 
we are deep in the Kondo regime since $T_{\rm{av}}=(T_L+T_R)/2 \ll T_K$ ($\approx \Gamma_0$).
In Fig.~\ref{rys5}(a) the charge current calculated for $T_L=T+\Delta T$, $T_R=T$ is compared to the  
one flowing in the system with the same temperature bias but applied to the right electrode. In both cases the 
base temperature of the device is the same and equals $T$, so the difference between the curves
of different colours directly demonstrates the rectification effect. Thin (thick) curves correspond to 
an asymmetry of 2 and 3, respectively. The current flow is in the direction consistent with temperature difference, 
and for sufficiently large values of $\Delta T$ their values are visibly different. The rectification 
factor amounts to about 30\% and depends on $\Delta T$. 

{Figure \ref{rys5}(b) shows a similar rectification of the heat current for the same model parameters and
the same asymmetry, $\Gamma_R/\Gamma_L=2,3$. However, for clarity
the curves calculated for $\Gamma_R/\Gamma_L=3$ have been shifted vertically by  $+2\cdot 10^{-3}$ for better visibility. 
The heat current rectification effect for this set of parameters is at least an order of magnitude smaller
than that for the charge and equals 1\% to 2\% in the parameter range shown in the figure. 
This is better visible in the inset which shows the $\Delta T >0.12\Gamma_0$ part of
the main figure. Calculating the charge and heat currents we have neglected the possible dependence of the
chemical potentials of the electrode with temperature.}

\section{Linear and non-linear thermopower and the role of asymmetry} \label{sec:nltep}

\begin{figure}
\begin{center}
\includegraphics[width=0.47\linewidth]{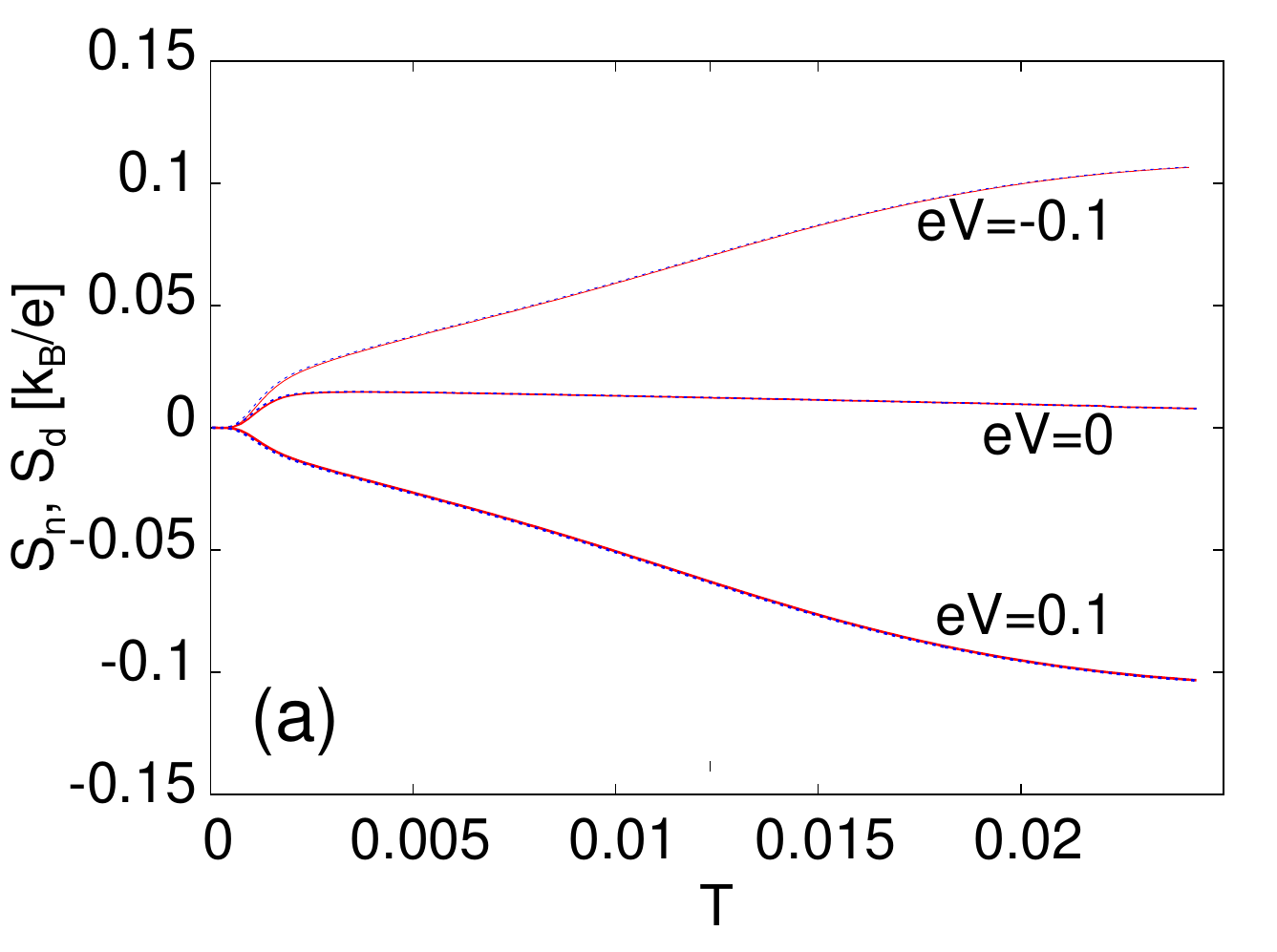}
\includegraphics[width=0.49\linewidth]{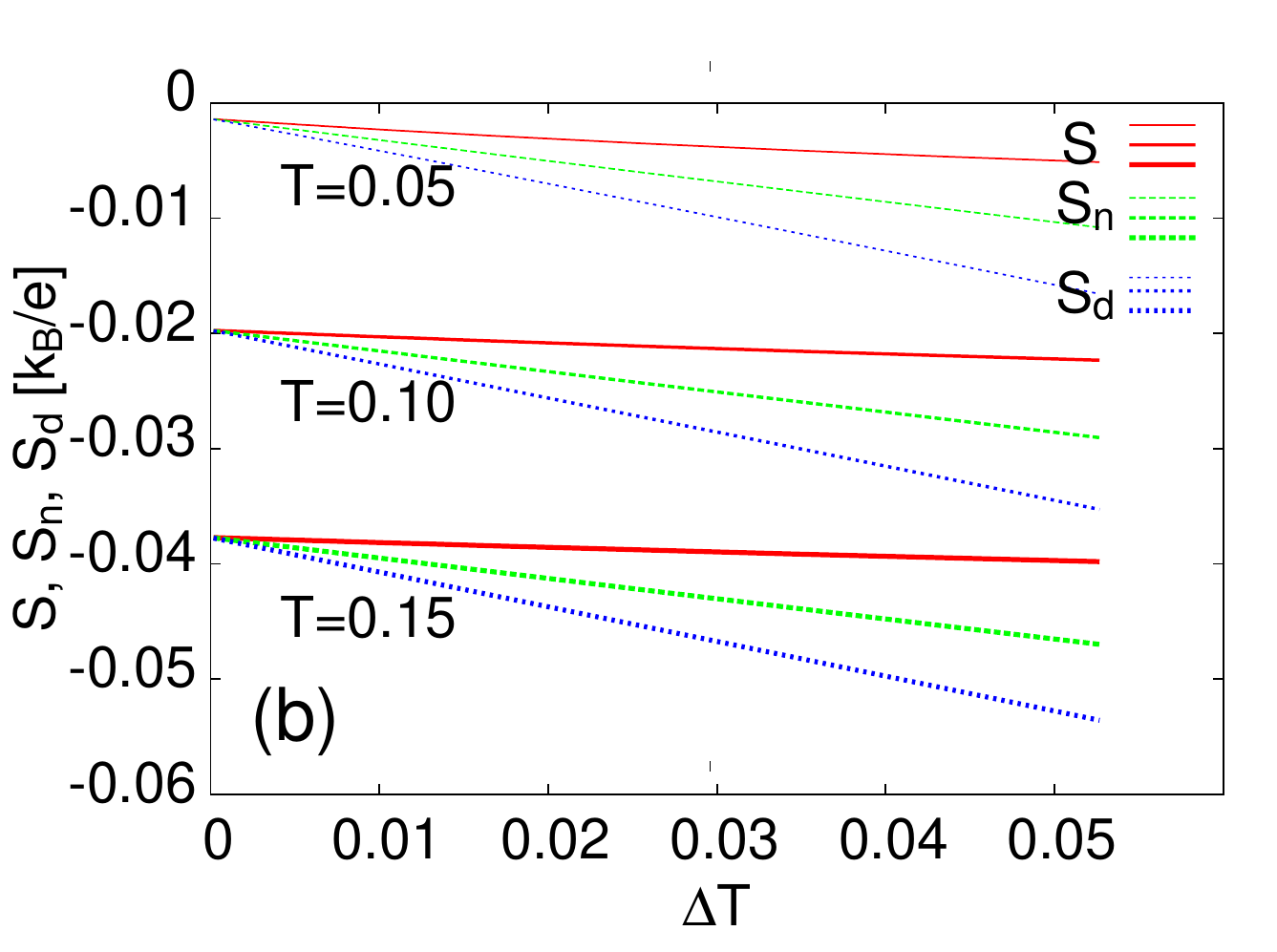}
\includegraphics[width=0.67\linewidth]{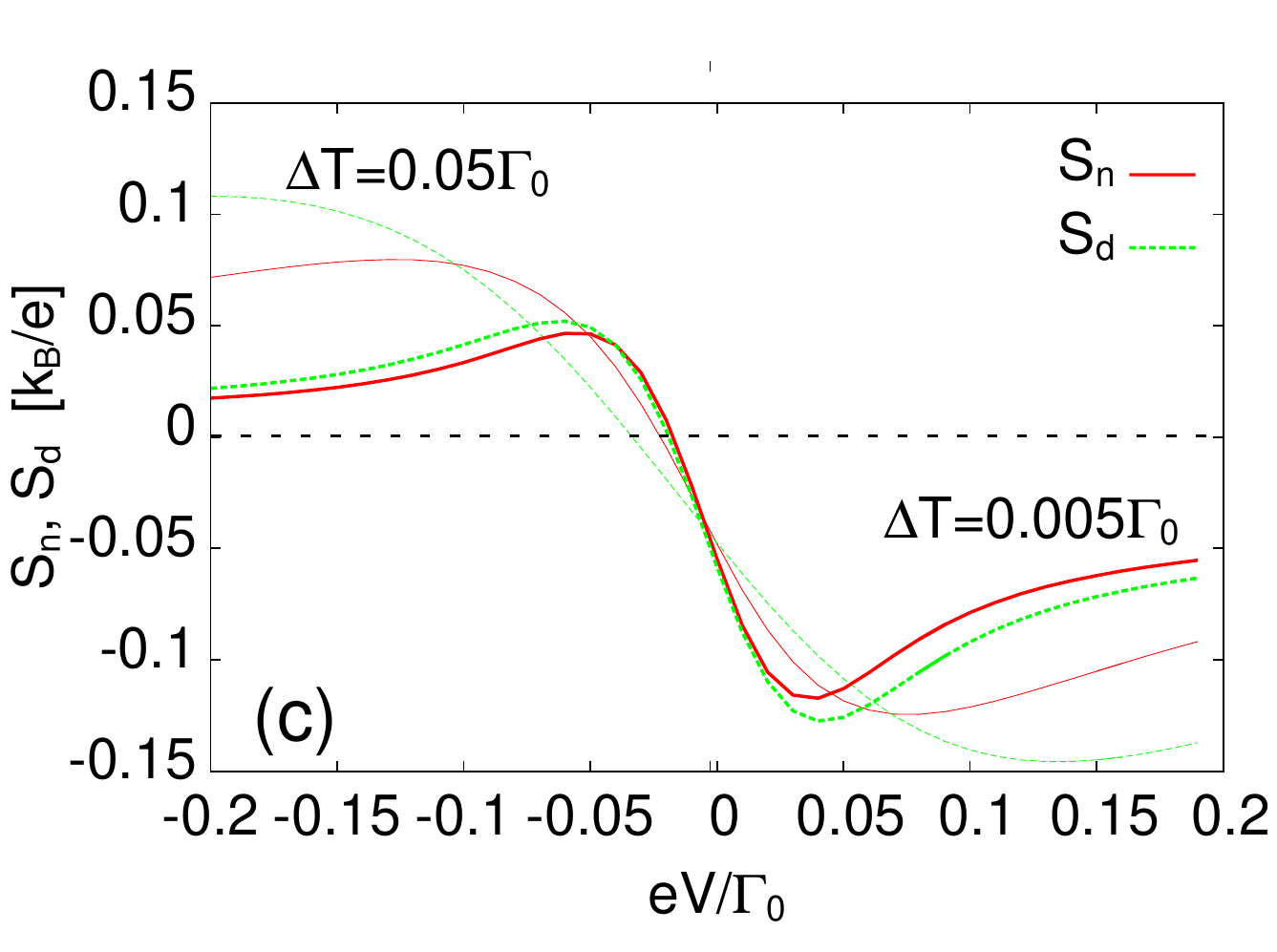}
\end{center}
\caption{(color online) (a) The middle curve shows 
that all three Seebeck coefficients (linear, $S$, non-linear, $S_n$, and differential, $S_d$)
of the two-terminal device calculated for $V=0$ and an infinitesimal value of $\Delta T$ 
agree with each other for all temperatures. The two other curves represent the non-linear 
thermopowers $S_n=S_d$, calculated for two values of bias, $eV=-0.1$ and $eV=0.1$. 
(b) Here we show the three thermopowers $vs.$
temperature difference $\Delta T$ between left and right electrodes, 
with $T_R=T$, $T_L=T+\Delta T$, calculated for  
the base temperatures $T=0.05$, $0.10$, and $0.15$, and for $V=0$. 
For large $\Delta T$'s strong non-linearities are clearly visible, as well as apparent differences between
the coefficients. (c) Voltage dependence of the non-linear
Seebeck coefficients $S_n$ and $S_d$ for two temperature differences, $\Delta T=0.005$ and $\Delta T=0.05$. 
Other parameters are: $U=12\Gamma_0$, $\varepsilon_d=-5\Gamma_0$, $\Gamma_0 \equiv \Gamma_L=\Gamma_R$.}
\label{rys6}
\end{figure}

Non-linear effects are expected to be important in all nano-devices whatever the
applied bias \cite{benenti2017,wysokinski2016}. The suitable
generalisation of the definition (\ref{seebeck}) of the Seebeck coefficient to non-linear situations reads
\begin{equation}
S_n=-\left(\frac{V}{\Delta T}\right)_{I(\Delta T, V)=0},
\label{seebeck-2}
\end{equation}
where $\Delta T$ is the temperature difference applied to the nanostructure. The traditional way to 
measure the Seebeck coefficient assumes the application of the temperature bias $\Delta T$,
and finding such  voltage $V$ where the current across the device vanishes, $I(\Delta T,V)=0$. 
In this formulation the only source of non-linearity is directly given by the value of $\Delta T$,
assumed to be large enough to preclude the linear expansion of the charge current $I(\Delta T,V)$ 
in first powers of $V$ and $\Delta T$. 

On the contrary, when the linear expansion is valid ($V \to 0$ and $\Delta T \to 0$), we have
\begin{equation}
I(\Delta T,V)=L_{11}V+L_{12}\Delta T,
\end{equation}
and the corresponding thermopower is given by the ratio of kinetic coefficients
\begin{equation}
S=\frac{L_{12}}{L_{11}}.
\label{seebeck-lin}
\end{equation}
Expanding Eq.~(\ref{curr-wbl}) for small $V$ and $\Delta T$ up to linear order, one finds an explicit
expression for $S$.

Sometimes the definition (\ref{seebeck-2}) of the non-linear Seebeck coefficient $S_n$ 
is extended to the differential one, $S_d$, 
calculated formally for constant current flowing as a result of the external voltage $V$.
This $S_d$ measures the response of the system with current flow to the 
change in temperature difference $\partial\Delta T$. At the applied external voltage $V$ and temperature
difference $\Delta T$, $S_d$ is defined \cite{dorda2016,daroca2018} as the derivative 
\begin{equation}
S_d=-\left(\frac{\partial V}{\partial \Delta T}\right)_{I=\mathrm{const}} 
=-\left(\frac{\partial I}{\partial \Delta T}\right)\bigg/\left(\frac{\partial I}{\partial V}\right).
\label{seebeck-dif}
\end{equation}
As argued earlier, $S_d$ should also be accessible experimentally \cite{dorda2016}.
In that paper the response $\left({\partial I}/{\partial \Delta T}\right)$  
has been calculated under the additional approximation that, at a given but arbitrary external voltage, 
a very small temperature bias $\Delta T$ is applied to the
system. 
{The definition (\ref{seebeck-dif}) can be viewed as analogous to the differential conductance 
$G(V)={\partial I}/{\partial V}$ often used to characterise the conductance in the non-linear regime.
Similarly to $G(V)$ also $S_d$ should be accessible experimentally in appropriate ac circuits.
We remark that $S_d$ has been analysed in Refs. \cite{dorda2016,daroca2018}, and the authors claim   
that this non-linear Seebeck coefficient allows for a better understanding of decoherence processes at
finite voltage. It may have potential applications in nanoscale temperature sensors.}

Here we are interested in the generalisation of $S_d$ to arbitrary temperature differences $\Delta T$ 
and arbitrary bias as well as its comparison to $S_n$ and $S$.  
Of course, for infinitesimally small $\Delta T$ and $V$ all definitions 
of Seebeck coefficients are equivalent and lead to the same result, $S=S_n=S_d$.
For arbitrary $V$ and $T$ but vanishingly 
small $\Delta T$, the two non-linear Seebeck coefficients are equal, $S_n=S_d$. 
The formula (\ref{seebeck-dif}) extends the standard definition (\ref{seebeck-2}) towards the
non-linear regime, caused by both a large $\Delta T$ and additionally a finite (large) 
externally applied bias voltage $V$. In this regime, $S_n$ and $S_d$ attain different values.

To gain some feeling about the relative values and the behaviour of the three, in principle different 
Seebeck coefficients we show in Fig.~\ref{rys6} the  
linear $S$ (where appropriate), non-linear $S_n$, and differential $S_d$ coefficients as
calculated for the two-terminal device.
The three panels in Fig.~\ref{rys6} illustrate their dependence on temperature $T$, temperature difference 
$\Delta T$, and voltage  $V$. We assumed $T_R=T$, $T_L=T+\Delta T$, $\mu_L=\mu+eV/2$, and $\mu_R=\mu-eV/2$,
and performed the calculations for $U=8\Gamma_0$, $\varepsilon_d=-5\Gamma_0$, with other parameters 
as given in the figure. We remark that for these parameters the Haldane formula for the Kondo temperature
(with $(\Gamma_L+\Gamma_R)/2=\Gamma_N$),
\begin{equation}
T_K=\frac{\sqrt{U\Gamma_N}}{2}\exp{\frac{\pi\varepsilon_d(\varepsilon_d+U)}{\Gamma_NU}},
\end{equation}
leads to $T_K\approx 0.071 \Gamma_0$.

Figure \ref{rys6}(a) illustrates the $T$-dependence of all Seebeck 
coefficients for three values of the voltage $eV=-0.1$, $0$, and $0.1$, and for a very small temperature 
difference, $\Delta T\rightarrow 0$. The value $V=0$ in fact denotes a very small
voltage, $V\rightarrow 0$. All parameters are given again in energy units of $\Gamma_0$. 
For the actual calculations, we have used $\delta T=10^{-9}$,
and to calculate the voltage derivative in (\ref{seebeck-dif}) we have used 
$\delta V=10^{-9}$.  All coefficients have the same value $S=S_n=S_d$ if $V=0$. 
For a relatively large value of $V$ the calculation of $S$ is meaningless; 
the figure shows that, independent of $T$ and for
both values of $V$ one has $S_n=S_d$, the equality being due to the smallness 
of the applied temperature difference.

The situation is different if $\Delta T$ is arbitrary. 
The  results for $V = 0$ are presented in Fig.~\ref{rys6}(b); it is apparent 
that all coefficients assume different values, except for small $\Delta T$ (the linear case). 
The differences increase with increasing $\Delta T$, with $S_d$ lying {below} $S_n$ (and $S_n$ {below} $S$) 
for all $T$ {and a given set of parameters}. 

For non-zero voltages the relative order of $S_n$ and $S_d$ may be different, as is visible from the
panel (c) of the figure. {The Seebeck coefficient $S_n$ provides a generalisation of the 
standard definition to the non-linear regime. On the other hand, the differential Seebeck coefficient
$S_d$ characterises the response to the temperature change of the voltage 
biased system.} The data presented in Fig.~\ref{rys6}(c) have been obtained for $T=0.01$, and two values
$\Delta T=0.005$ and $0.05$. For small temperature bias the curves 
corresponding to $S_n$ and $S_d$ are rather close to each other. However, for larger $\Delta T$
the non-linear dependence of the Fermi functions and the on-dot density of states on $V$ and $\Delta T$ 
lead to various contributions to both $S_n$ and $S_d$.

The non-zero value of both Seebeck coefficients $S_n$ and $S_d$ for $V=0$ can be 
understood by taking into account the lack of 
particle-hole symmetry, $2\varepsilon_d+U\ne 0$, for the set of parameters used. For these 
parameters the density of states is similar to that shown in the panel (a) of Fig.~\ref{rys2}.
The differences between the curves $S_n(V)$ and $S_d(V)$ for the same $\Delta T$ are of the same character 
as those observed in the middle panel of the figure, while the global similarities 
between the two sets of curves 
calculated for $\Delta T=0.005$ and $\Delta T=0.05$ can be traced back to a larger contribution to  
$(\partial I / \partial \Delta T)$ obtained from Eq.~(\ref{curr-wbl}): 
\begin{eqnarray}
\left(\frac{\partial I}{\partial \Delta T}\right) 
& \approx & - \frac{2e}{\hbar} \sum_\sigma \tilde{\Gamma}_\sigma \int\frac{dE}{2}
\bigg\{ \left[f_L(E,T)-f_R(E,T)\right] \left(\frac{\partial N(E,T,\Delta T)}{\partial \Delta T}\right) \nonumber \\
& - & N(E,T,\Delta T)\left(-\frac{\partial f_L(E)}{\partial \Delta T}\right) \bigg\} +  O((\Delta T)^2) +\cdots
\label{curr-wbl-dT-exp}
\end{eqnarray}
Note that the difference
\begin{equation}
D(V)=[f_L(E,T)-f_R(E,T)]=-D(-V)
\end{equation}
is an odd function of the voltage, which implies that the resulting curves are nearly anti-symmetric with respect to $V=0$ 
(and that the ordinates are equal, $S_n(0)\approx S_d(0)$). The smaller contribution proportional to  
$\left(-{\partial f_L(E)}/{\partial \Delta T}\right)$ depends on $V$ in a non-universal way. This causes the 
crossing of two coefficients at various voltages and for $\Delta T$ well beyond the validity of the linear approximation.

\begin{figure}
\centerline{\includegraphics[width=0.65\linewidth]{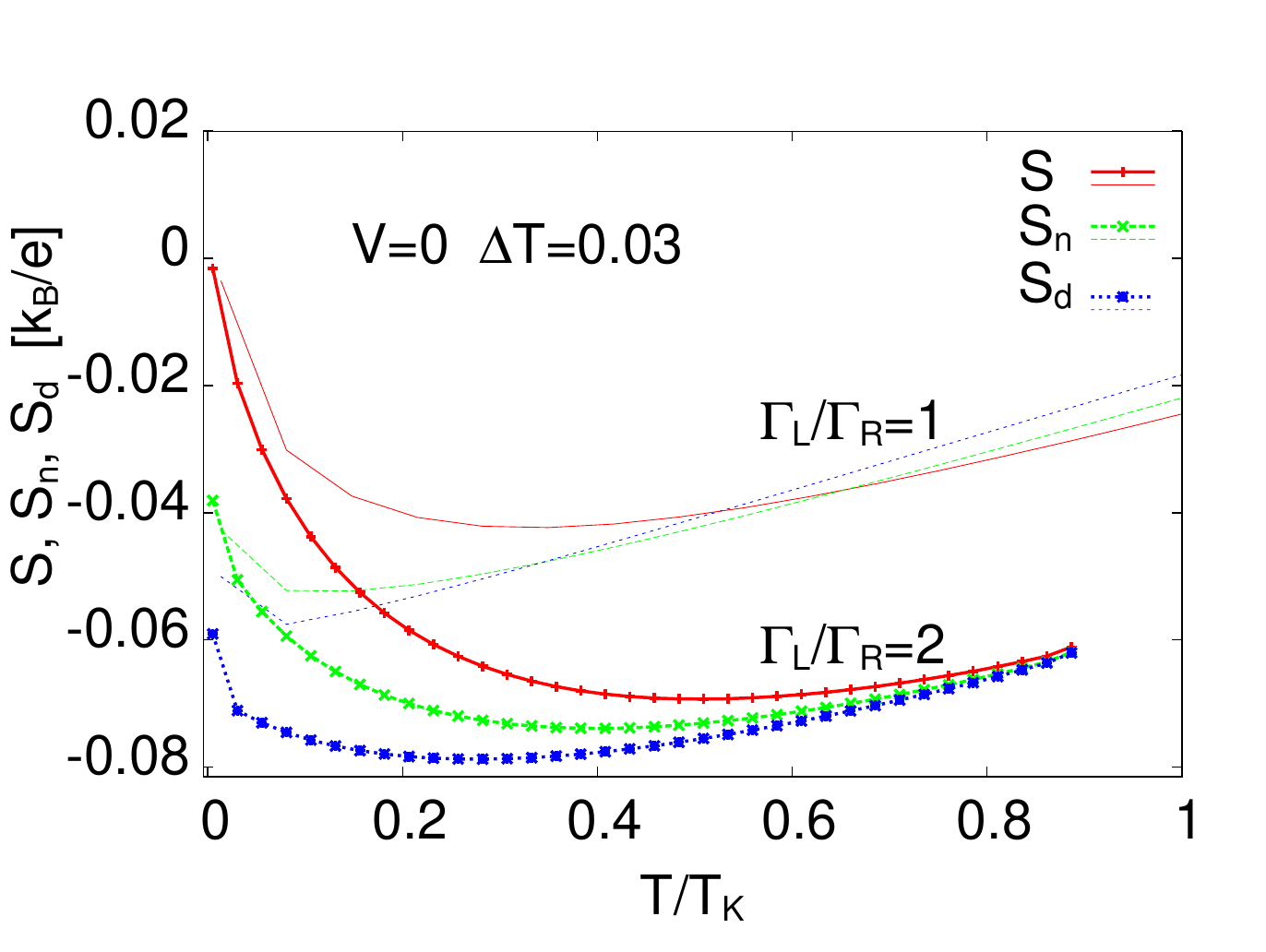}}
\caption{(color online) The anisotropy of the couplings $\Gamma_R/\Gamma_L\neq 1$ results in 
increased absolute values of the Seebeck coefficients. We show all three coefficients, albeit
the temperature difference $\Delta T=0.03\Gamma_0$ is slightly beyond the validity of the linear approximation. 
Other parameters of the model read: $U=8\Gamma_0$, $\varepsilon_d=-5\Gamma_0$.
The Kondo temperature is $T_K\approx 0.004\Gamma_0$ for the symmetric case, $\Gamma_R/\Gamma_L=1$, and 
$\approx 0.034\Gamma_0$ for $\Gamma_R/\Gamma_L=2$.}
\label{rys7}
\end{figure}

As a final remark we note  that the asymmetry of the couplings $\Gamma_L \neq \Gamma_R$  
also affects the Seebeck coefficients.
To see this we assume $V=0$ and a relatively small (but slightly beyond 
the validity of the linear approximation) temperature difference $\Delta T=0.03$.  
Figure~\ref{rys7} shows the three Seebeck coefficients $S$, $S_n$, and $S_d$, calculated for isotropic coupling 
$\Gamma_R/\Gamma_L=1$ (thin lines) as well as for an anisotropic system $\Gamma_R/\Gamma_L=2$
(thick lines). Interestingly, the largest decrease of the Seebeck coefficient 
is observed at low temperatures, well below the Kondo temperature $T_K$, 
in agreement with recent findings \cite{daroca2018} based on the non-crossing approximation. 
These authors already noted that the effect is largest in the Kondo regime. The
asymmetry of the couplings is an important experimental fact \cite{delagrange2018} which has 
to be taken into account in any calculation aiming at a comparison with experiment \cite{crepieux2018}. 
Experimentally, it has  been found \cite{delagrange2018} that, $e.g.$, the asymmetry shifts the cut-off of the noise 
emission from quantum dots from lower towards higher frequencies.

\section{Three-terminal heat engine: the role of strong Coulomb interactions} \label{sec:nlheat}
{As already noted a thermoelectric device can serve as waste heat to electricity 
converter. The energy harvesting is an important issue at large and small scales. As discussed in the Introduction, 
the usefulness of the bulk
material as heat engine is well characterised by the figure of merit, $ZT$. This parameter ceases to be a good 
quality indicator in nanostructures \cite{muralidharan2012,wysokinski2016}, and the calculations of both 
the power and the efficiency are required. The quantity of interest then is the
maximum power, and the efficiency at maximum power.}

In this section we consider the effects of strong interactions between 
the on-dot electrons on the characteristics of
the three-terminal heat engine consisting of two cold terminals connected 
{\it via} two quantum dots to the hot one, cf.\ Fig.~\ref{rys1}.
Such a three-terminal heat engine, for the non-interacting case but well outside equilibrium, 
has been studied earlier \cite{jordan2013,donsa2014}.
The work has been later extended \cite{szukiewicz2016} to take screening effects into account,
treating the interactions within a mean-field approximation. The main conclusion was that 
albeit the screening modifies the parameters at which the engine is optimal, it does neither change 
the maximal power nor the efficiency at the maximum power. A similar heat engine has also been 
optimized with respect to the transmission function \cite{schiegg2017}.

Here we shall pursue the analysis assuming non-linear working conditions and  
taking strong interactions into account. {The calculations include the Kondo regime.}  
To  this end, we use the general expressions (\ref{charge-curr}) for 
the charge and (\ref{heat-curr}) for the heat current flowing out of the $\lambda$ lead, 
the energy conservation (\ref{en-cons}), and the general expression (\ref{sol-gf}) for the on-dot Green function.    
From the charge current flowing from the left to the right electrode, 
and the heat current flowing from the hot to the cold electrodes, we
calculate the performance of the engine as quantified by the maximum power $P$ and the 
efficiency $\eta$ at maximum power. 

Previously we have found \cite{szukiewicz2016} that the three-terminal heat engine
at optimized conditions attains an efficiency slightly above 0.2 in units of the Carnot efficiency,
and that this value is roughly the same as without screening effects. 
The optimization involved the effective coupling between the dots and the leads, 
$\Gamma/T_{\rm{av}}$, the temperature difference 
between the electrodes, $\Delta T/T_{\rm{av}}$, and the load voltage, $V$; $T_{\rm{av}}$ is the average
temperature of the system. The calculations have shown that the optimal coupling, {{\it i.e.}, 
the one leading to the maximal power if other parameters remain fixed,} is of the order of
the average temperature, $\Gamma/T_{\rm{av}}\approx 1$. (The coupling $\Gamma$ introduced here
refers to the symmetric engine with $\Gamma_L=\Gamma_R =\Gamma_H \equiv\Gamma$.)
The efficiency has been found to depend rather weakly,  
and the power strongly on the temperature difference $\Delta T$ between the hot ($H$), 
$T_H=T+\Delta T$, and the two cold ($R,L$) electrodes, $T_R=T_L=T$.
 
\begin{figure}
\begin{center}
\includegraphics[width=0.47\linewidth]{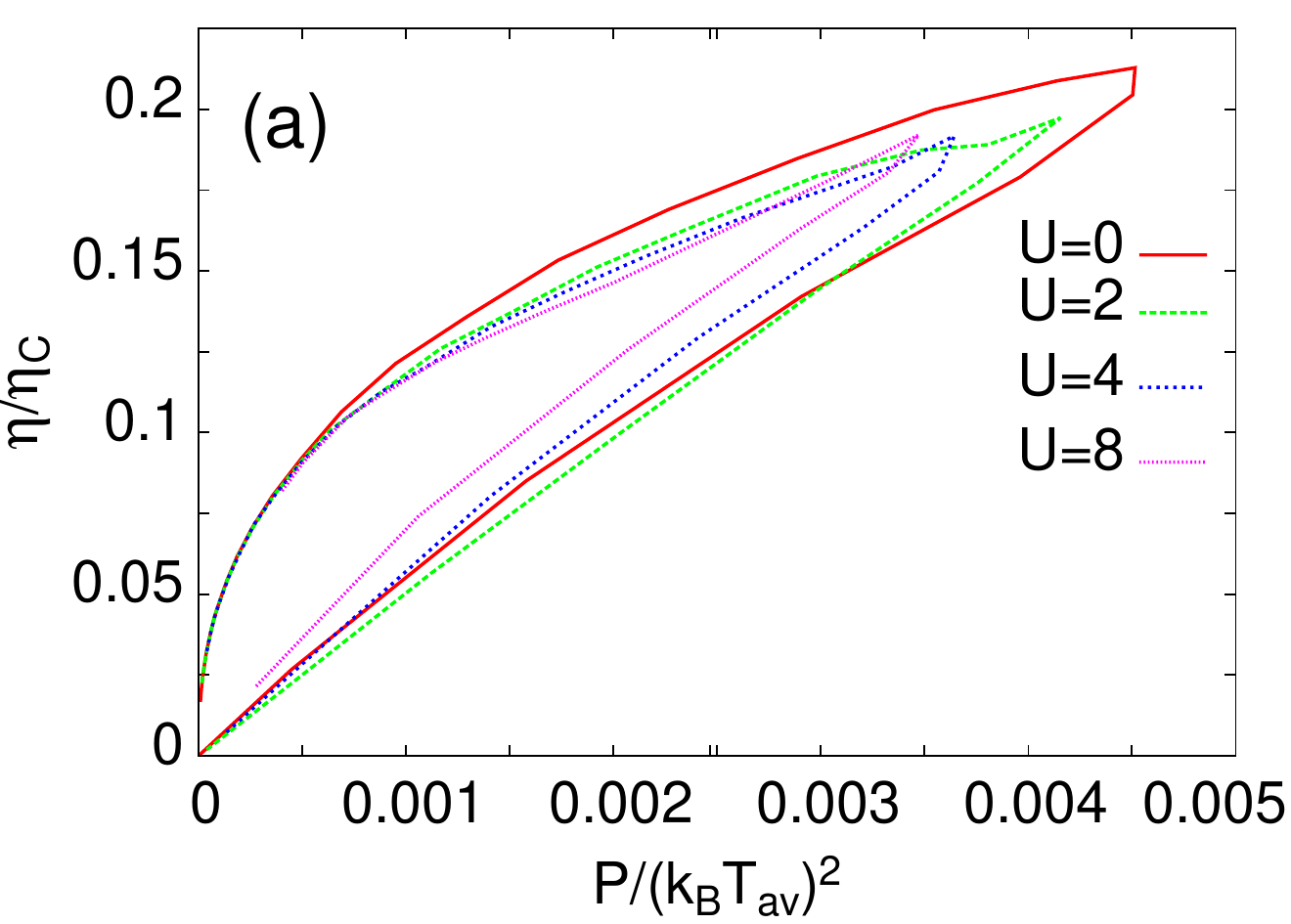}
\includegraphics[width=0.47\linewidth]{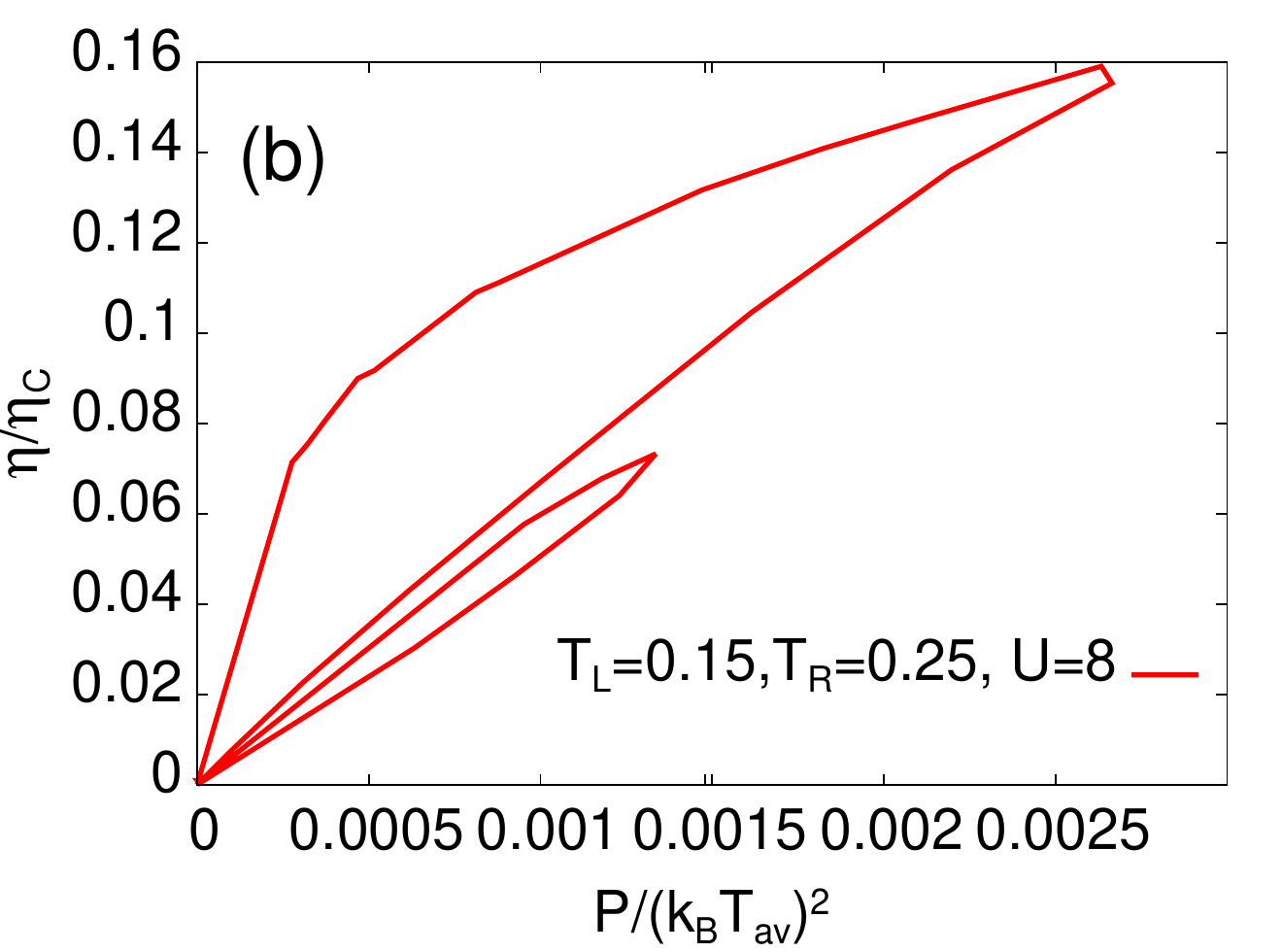}
\end{center}
\caption{(color online) Panel (a) shows efficiency $vs.$ power for the three-terminal quantum dot engine,
assuming optimal values of the couplings:
$\Gamma_L=\Gamma_R=\Gamma_H\equiv \Gamma=T_{\rm{av}}$, with $\Delta T=0.5$, $T_L=T_R=T=0.75$, $T_H=T+\Delta T=1.25$, 
and a few values of the Coulomb interaction parameters $U$ (energies in units of $\Gamma_0 = \Gamma$).
Panel (b) shows the efficiency $vs.$ power plot for a two-terminal system. For $U=8\Gamma_0$ 
and a range of values of the dot energies $\varepsilon_d$, the Kondo effect is observed at the considered 
temperatures. The Kondo effect results in the second (lower) branch
on the $\eta$-$P$ plane. In the Kondo region the performance of the system as 
a heat engine does not exceed that in the other region.}
\label{rys8}
\end{figure}

In order to demonstrate the effect of the Coulomb interaction on the performance of the engine, we show 
in  Fig.~\ref{rys8} the efficiency $\eta$ $vs.$ power $P$. The efficiency is measured in units of the  
Carnot efficiency, $\eta_C={\Delta T}/{T_R}=0.4$, and
the power is normalized by $(k_B T_{\rm{av}})^2$. The plot shown in panel (a) of the figure, 
valid for a three-terminal system, has been obtained 
by calculating the heat and charge currents as well as the optimal voltage (and the power) for a given difference 
of the dot's energy levels, $\Delta E=\varepsilon_R-\varepsilon_L$, where $\varepsilon_R$  
($\varepsilon_L$) refers to the energy level of the right (left) dot in the system (see Fig.~\ref{rys1}). 
It has to be noted that for appropriate values of $\varepsilon_R$  ($\varepsilon_L$) and low temperature 
the respective  dot may show Kondo behaviour. This has an important effect on the $\eta$--$P$ plot. 

In panel (a) of Fig.~\ref{rys8} one sees that the $U=0$ curve 
essentially encompasses all curves obtained for the interacting case. Coulomb interactions
generally suppress the performance characteristics -- at least so under the assumptions 
of the present approach, including the wide band approximation. The temperature of
all electrodes are such that this system (Fig.~\ref{rys8}(a)) is always outside the Kondo regime.

To illustrate the behaviour of the engine working  
in the Kondo regime, we show in  Fig.~\ref{rys8}(b) a similar plot, but obtained 
for a two-terminal system and a much lower value of the average temperature. For the interaction $U=8\Gamma_0$ and   
temperatures $T_L=0.15\Gamma_0$, $T_R=0.25\Gamma_0$, the dot enters 
the Kondo regime for a range of gate voltages (or $\varepsilon_d$). The  Kondo 
effect existing for some values of $\varepsilon_d$ results in the appearance of the new performance branch  
on the $\eta$--$P$ plane. As visible in the  panel (b) of the figure, this region is characterized by low
efficiencies and powers when the system works as a heat engine.

The three-terminal quantum dot working as a refrigerator at very low temperature and in the Kondo regime 
requires a more careful examination, which is outside the scope of the present paper.

The results shown in Fig.~\ref{rys8} have been obtained under the assumption that the couplings to the leads 
equals the average temperature, {$\Gamma=k_BT_{\rm{av}}$}, the value which has been found to be optimal ({\it i.e.}, 
leading to the maximal power) for the non-interacting system.
{This facilitates} the comparison with the previous results \cite{szukiewicz2016}. 
The red curve in Fig.~\ref{rys8}(a), corresponding to the non-interacting case, agrees with previously obtained 
results for the same set of parameters: $\Gamma_L=\Gamma_R=\Gamma_H=T$, $\Delta T/T_{\rm{av}}=0.5$. 
Analogous to Ref. \cite{szukiewicz2016}, 
{changing the dots energy levels we fix the difference $\Delta E=\varepsilon_R-\varepsilon_L$ and calculate} 
the power and efficiency at each point for the optimized value of the load voltage $V$. 
{The curves in Fig.~\ref{rys8} are thus parametrised by $\Delta E$.}
As is apparent from the figure, the maximal efficiency is slightly higher than 0.2$\eta_C$, 
{which nicely agrees with recent measurements on similar systems \cite{jaliel2019}.  }

\section{Summary and conclusions}\label{sec:sum}

We have studied the thermoelectric transport properties of  two- and three-terminal systems with quantum dots, 
paying attention to strong interactions of electrons on the dot(s) and far-from-equilibrium conditions. 
Of particular importance in such nanostructures is the non-linear regime at large external voltage bias $V$ and 
large temperature difference $\Delta T$. As discussed in the Introduction, the linear approximation is hardly ever valid
in nanostructures. This has been again confirmed here, and is visible as the detrimental effect of strong correlations 
on the performance of the three-terminal optimized heat engine. Calculating the maximum power $P$ and the efficiency
$\eta$ of the optimized heat engine for various interactions $U$, we have observed that except at very low temperatures 
the curves calculated for $U\ne 0$ are encompassed by the curve obtained for the non-interacting system.
Obviously the filtering properties are affected by the interactions which broaden or split the density 
of states rendering the filter less effective. This agrees with previous attempts at optimizing heat engines  
by engineering the transmission function \cite{hershfield2013,schiegg2017}. On the other hand,
strong interactions are responsible for the second leaf on the efficiency {\it vs.} power plane,
{visible in Fig.~\ref{rys8}(b) and} appearing 
at low temperatures when the system enters the Kondo regime, {albeit for a relatively small} 
range of gate voltages. This novel feature {appearing in the Kondo regime} requires 
further analysis, especially for a three-terminal quantum dot working as a refrigerator {at very low temperatures}.      

The strongly non-linear regime requires proper definitions of the thermopower. 
The generalisation of the standard linear-response definition of the Seebeck coefficient 
to the strongly non-linear regime has been considered. This results in two different formulas for 
two coefficients $S_n$ and $S_d$. Both are, in principle, valid for arbitrary values of $V$ and $\Delta T$,
albeit the first coefficient ($S_n$) is easier to handle in systems with zero external voltage but arbitrarily large 
temperature difference between a given pair of electrodes. The second, called differential Seebeck coefficient,
has earlier been applied \cite{dorda2016,daroca2018} to systems with finite current flow and a small temperature 
difference. It has been generalized and studied here for arbitrary $\Delta T$ in the presence of an external 
voltage bias $V$. Interestingly the asymmetry of the couplings to the external leads has qualitatively similar 
influence on both Seebeck coefficients. They develop a minimum for temperatures well below the Kondo temperature. 
The observed quantitative differences between $S_n$ and $S_d$ are expected to be important for temperature 
sensing by means of thermopower measurements \cite{kleinbaum2018}. 
{The non-symmetric device with different couplings between left and right electrode can, in principle, 
work as a thermal diode, albeit with a rather small heat rectification factor of the order of 1\%. 
The analogous rectification factor  for a charge flow is more than an order of magnitude higher, up to about 30\%.}

\ack{The work reported here has been supported by the M.~Curie-Sk\l{}odowska University,
National Science Center grant DEC-2017/27/B/ST3/01911 (Poland),
the Deutsche Forschungsgemeinschaft (DFG, project no.\ 107745057, TRR 80), and the University of Augsburg.}

\appendix
\section{Details of the calculations}\label{sec:equ}
There exist many techniques which provide a correct description of the physics
leading to the single-impurity Kondo effect \cite{hewson}. However, most of them are based
on numerical approaches very often restricted to equilibrium situations.
A quantitatively correct analytical description of the Kondo effect valid both in and out of equilibrium,
however, is important as it provides additional insights into the underlying physics \cite{smirnov2013,galperin2007}.
Here we shall show that the appropriately generalized \cite{lavagna2015} EOM technique supplemented with
an analytic calculation of the lifetimes provides an easy, physically transparent, and numerically correct
approach to study the Kondo regime.

To calculate the retarded GF in the stationary state, it is convenient to use the EOM 
for two-time functions \cite{zubarev1960} in the energy representation.  
The equation of motion for an arbitrary Green function 
$\langle \langle A|B\rangle\rangle_E$ composed of the fermionic operators $A$ and $B$ reads:
\begin{equation}
E\langle \langle A|B\rangle\rangle_E=\langle \{A,B\}\rangle +\langle \langle [A,H]|B\rangle\rangle_E
\label{eom}
\end{equation}
However, the calculation of lesser 
Green functions are more conveniently performed on the time contour and continued to real times 
by means of Langreth's theorem \cite{langreth1976}. This approach has been used to obtain the average 
number of electrons on the dot, Eq.~(\ref{n-noneq}).

To determine the retarded GF, one starts with the equation of motion  
for $g^r_\sigma(E)=\langle\langle d_\sigma|d^\dagger_\sigma\rangle\rangle_{E+i0}$, which reads
\begin{equation}
E \langle \langle  d_\sigma|d^\dagger_\sigma\rangle\rangle_E=\langle  \{d_\sigma,d^\dagger_\sigma\}\rangle
+\langle \langle [d_\sigma,H]|d^\dagger_\sigma\rangle\rangle_E,
\label{eom-g}
\end{equation}
where $H$ denotes the Hamiltonian (\ref{ham1}) of the system.

Calculating the commutators one finds
\begin{equation}
[E-\varepsilon_{\sigma}]\langle\langle d_{\sigma}|d^{\dagger}_{\sigma}\rangle\rangle_{E} = 
1 + \sum_{\lambda k}V^{*}_{\lambda k \sigma}\langle\langle c_{\lambda k \sigma} |d^{\dagger}_{\sigma}\rangle\rangle_{E} 
 +  U\langle\langle n_{\bar{\sigma}}d_{\sigma}|d^{\dagger}_{\sigma}\rangle\rangle_{E}. 
\label{(A-g)}
\end{equation}
Here the symbol $\bar{\sigma} \equiv -\sigma$ is introduced, and  
two new GFs show up on the r.h.s. In turn, we calculate both of them, 
again using the equation of motion (\ref{eom}). The simplest one is given by
\begin{eqnarray}
(E-\varepsilon_{\lambda k})\langle\langle c_{\lambda k \sigma} |d^{\dagger}_{\sigma}\rangle\rangle_{E}
=V_{\lambda k \sigma}\langle\langle d_{\sigma}|d^{\dagger}_{\sigma}\rangle\rangle_{E}.
\label{eq:D3}
\end{eqnarray}
In Eq.~(\ref{(A-g)}) we need this function multiplied by $V^{*}_{\lambda k \sigma}$ and 
summed over $\lambda k$. The result is: 
\begin{equation}
\sum_{\lambda k}V^{*}_{\lambda k \sigma}\langle\langle c_{\lambda k \sigma} |d^{\dagger}_{\sigma}\rangle\rangle_{E}=
\sum_{\lambda k}\frac{|V_{\lambda k \sigma}|^2}{E+i0-\varepsilon_{\lambda k}} \langle\langle d_{\sigma}|d^{\dagger}_{\sigma}\rangle\rangle_{E}.
\label{Dd}
\end{equation}
The factor in front of the GF on the r.h.s.\ defines the self-energy:
\begin{eqnarray}
\Sigma_{0\sigma} (E)=\sum_{\lambda k} \frac{|V_{\lambda k \sigma}|^{2}}{E+i0-\varepsilon_{\lambda k}}.
\label{sigma0}
\end{eqnarray}
In the wide band limit, one approximates (\ref{sigma0}) by its imaginary part:
\begin{equation}
\Sigma_{0\sigma} (E)\approx -i\pi \sum_{\lambda k}{|V_{\lambda k \sigma}|^{2}}\delta(E-\varepsilon_{\lambda k}) 
=-i\frac{1}{2}\sum_\lambda \Gamma_\sigma^\lambda(E) 
=-i\bar{\Gamma}_\sigma/2,
\label{Gam-wbl}
\end{equation}
and  neglects the possible energy dependence of $\Gamma_\sigma^\lambda(E)=\Gamma_\sigma^\lambda$.
As discussed in Sec.~\ref{sec:curr}, the wide band approximation is essential in 
the derivation of the exact formula (\ref{n-noneq}).

In principle it is tempting to decouple the GF on the r.h.s.\ of (\ref{(A-g)}) multiplying $U$ 
as $\langle\langle n_{\bar{\sigma}} 
d_{\sigma}|d^{\dagger}_{\sigma}\rangle\rangle _{E}\approx \langle n_{\bar{\sigma}}\rangle\langle\langle  
d_{\sigma}|d^{\dagger}_{\sigma}\rangle\rangle _{E}$, but it turns out that the equation of motion for this
GF introduces important new functions, describing fluctuations of opposite-spin ($\bar{\sigma}$) electrons.
In order to obtain a qualitatively correct description of Kondo correlations, this function has to be calculated 
exactly \cite{lacroix1981,sim-appr} at this level. In the next step, we obtain:  
\begin{eqnarray}
[E-\varepsilon_{\sigma}-U]\langle\langle n_{\bar{\sigma}}d_{\sigma} |d^{\dagger}_{\sigma}\rangle\rangle _{E} 
= \langle n_{\bar{\sigma}}\rangle 
- \sum_{\lambda k} V_{\lambda k \bar{\sigma}}\langle\langle c^{\dagger}_{\lambda k \bar{\sigma}} d_{\bar{\sigma}} 
d_{\sigma} |d^{\dagger}_{\sigma}\rangle\rangle_{E} 
\nonumber \\ 
+  \sum_{\lambda k} \big[ V^{*}_{\lambda k\sigma}\langle\langle n_{\bar{\sigma}} c_{\lambda k \sigma} 
|d^{\dagger}_{\sigma}\rangle\rangle_{E}   
+  V^{*}_{\lambda k \bar{\sigma}}\langle\langle d^{\dagger}_{\bar{\sigma}} c_{\lambda k \bar{\sigma}} d_{\sigma} | 
d^{\dagger}_{\sigma}\rangle\rangle_{E} \big]
\label{(D-g)}
\end{eqnarray}
Interestingly, the GFs containing one lead operator enter the above equations {\it via} the sums
\begin{eqnarray}
S_n &=& \sum_{\lambda k}V^{*}_{\lambda k \sigma}\langle\langle n_{\bar{\sigma}} c_{\lambda k \sigma} |d^{\dagger}_{\sigma}\rangle\rangle_{E} , \\
S_d &=& \sum_{\lambda k}V_{\lambda k \bar{\sigma}}\langle\langle c^{\dagger}_{\lambda k \bar{\sigma}} d_{\bar{\sigma}} 
d_{\sigma} | d^{\dagger}_{\sigma}\rangle\rangle_{E} , \\
S_c &=& \sum_{\lambda k}V^{*}_{\lambda k \bar{\sigma}}\langle\langle d^{\dagger}_{\bar{\sigma}} c_{\lambda k\bar{\sigma}} d_{\sigma} |d^{\dagger}_{\sigma}\rangle\rangle_{E},
\end{eqnarray}
which are calculated using again the EOM for the new GFs. They read 
\begin{eqnarray}
[E-\varepsilon_{\lambda k}]\langle\langle n_{\bar{\sigma}}c_{\lambda k\sigma}|d^{\dagger}_{\sigma}\rangle\rangle_{E} 
 = V_{\lambda k\sigma}\langle\langle n_{\bar{\sigma}} d_{\sigma}|d^{\dagger}_{\sigma}\rangle\rangle_{E} 
- \sum_{\lambda' k'} V_{\lambda' k'\bar{\sigma}}\langle\langle c^{\dagger}_{\lambda' k'\bar{\sigma}}
 d_{\bar{\sigma}} c_{\lambda k \sigma}|d^{\dagger}_{\sigma}\rangle\rangle_{E}  \nonumber \\
+ \sum_{\lambda'k'}V^{*}_{\lambda'k'\bar{\sigma}}\langle\langle d^{\dagger}_{\bar{\sigma}} c_{\lambda' k'\bar{\sigma}}
c_{\lambda k \sigma}|d^{\dagger}_{\sigma}\rangle\rangle_{E},
\label{(B-g)}
\end{eqnarray}
\begin{eqnarray}
[E -\varepsilon_{\lambda k}-\varepsilon_{\sigma} +\varepsilon_{\bar{\sigma}}]\langle\langle 
d^{\dagger}_{\bar{\sigma}} c_{\lambda k \bar{\sigma}} d_{\sigma}| d^{\dagger}_{\sigma} \rangle\rangle _{E} = 
  \langle d^{\dagger}_{\bar{\sigma}} c_{\lambda k \bar{\sigma}}\rangle + 
V_{\lambda k \bar{\sigma}}\langle\langle n_{\bar{\sigma}} 
d_{\sigma}|d^{\dagger}_{\sigma}\rangle\rangle _{E}  \nonumber \\
- \sum_{\lambda'k'} V_{\lambda' k' \bar{\sigma}}\langle\langle 
c^{\dagger}_{\lambda' k' \bar{\sigma}}c_{\lambda k \bar{\sigma}} d_{\sigma}|d^{\dagger}_{\sigma}\rangle\rangle _{E} 
+ \sum_{\lambda'k'} V^{*}_{\lambda'k'\sigma}\langle\langle d^{\dagger}_{\bar{\sigma}} c_{\lambda k \bar{\sigma}} 
c_{\lambda'k'\sigma}|d^{\dagger}_{\sigma}\rangle\rangle _{E},
\label{(C-g)}
\end{eqnarray}
\begin{eqnarray}
[E+\varepsilon_{\lambda k}-\varepsilon_{\sigma}-\varepsilon_{\bar{\sigma}}
-U]\langle\langle c^{\dagger}_{\lambda k \bar{\sigma}} d_{\bar{\sigma}} 
d_{\sigma} | d^{\dagger}_{\sigma}\rangle\rangle_{E}= 
 \langle c^{\dagger}_{\lambda k \bar{\sigma}} d_{\bar{\sigma}}\rangle 
-V^{*}_{\lambda k \bar{\sigma}}\langle\langle n_{\bar{\sigma}} d_{\sigma} |d^{\dagger}_{\sigma}\rangle\rangle_{E} \nonumber \\
+ \sum_{\lambda' k'}\big[ V^{*}_{\lambda' k'\sigma}\langle\langle c^{\dagger}_{\lambda k \bar{\sigma}} d_{\bar{\sigma}} 
c_{\lambda' k' \sigma} |d^{\dagger}_{\sigma}\rangle\rangle_{E}  
+V^{*}_{\lambda' k' \bar{\sigma}} \langle\langle c^{\dagger}_{\lambda k \bar{\sigma}} c_{\lambda' k' \bar{\sigma}} 
d_{\sigma} |d^{\dagger}_{\sigma}\rangle\rangle _{E} \big]. 
\label{(E-g)}
\end{eqnarray}
To close the infinite hierarchy of consecutive equations, we have to employ an approximation at
a certain level. The common approximation is to project those new GFs appearing in the above set
which contain two lead operators onto the lower order ones, $e.g.$:
\begin{equation}
\langle\langle c^{\dagger}_{\lambda' k'\bar{\sigma}}  d_{\bar{\sigma}} c_{\lambda k \sigma}|d^{\dagger}_{\sigma}\rangle\rangle_{E} 
\approx \langle c^{\dagger}_{\lambda' k'\bar{\sigma}}  d_{\bar{\sigma}}\rangle \langle\langle c_{\lambda k \sigma}|d^{\dagger}_{\sigma}\rangle\rangle_{E} 
\end{equation}
\begin{equation}
\langle\langle d^{\dagger}_{\bar{\sigma}} c_{\lambda' k'\bar{\sigma}} c_{\lambda k \sigma}|d^{\dagger}_{\sigma}\rangle\rangle_{E} 
\approx \langle d^{\dagger}_{\bar{\sigma}} c_{\lambda' k'\bar{\sigma}} \rangle \langle\langle c_{\lambda k \sigma}|d^{\dagger}_{\sigma}\rangle\rangle_{E} 
\end{equation}
\begin{equation}
\langle\langle c^{\dagger}_{\lambda' k' \bar{\sigma}}c_{\lambda k \bar{\sigma}} d_{\sigma}|d^{\dagger}_{\sigma}\rangle\rangle _{E} 
\approx \langle c^{\dagger}_{\lambda' k' \bar{\sigma}}c_{\lambda k \bar{\sigma}}\rangle \langle\langle d_{\sigma}|d^{\dagger}_{\sigma}\rangle\rangle_{E} 
\end{equation}
\begin{equation} 
\langle\langle d^{\dagger}_{\bar{\sigma}} c_{\lambda k \bar{\sigma}} c_{\lambda'k'\sigma}|d^{\dagger}_{\sigma}\rangle\rangle _{E} 
\approx \langle d^{\dagger}_{\bar{\sigma}} c_{\lambda k \bar{\sigma}}\rangle \langle\langle c_{\lambda'k'\sigma}|d^{\dagger}_{\sigma}\rangle\rangle _{E} 
\end{equation}
\begin{equation}
\langle\langle c^{\dagger}_{\lambda k \bar{\sigma}} d_{\bar{\sigma}} c_{\lambda' k' \sigma} |d^{\dagger}_{\sigma}\rangle\rangle_{E} 
\approx \langle c^{\dagger}_{\lambda k \bar{\sigma}} d_{\bar{\sigma}} \rangle \langle\langle c_{\lambda' k' \sigma} |d^{\dagger}_{\sigma}\rangle\rangle_{E} 
\end{equation}
\begin{equation}
\langle\langle c^{\dagger}_{\lambda k \bar{\sigma}} c_{\lambda' k' \bar{\sigma}} d_{\sigma} |d^{\dagger}_{\sigma}\rangle\rangle _{E} 
\approx \langle c^{\dagger}_{\lambda k \bar{\sigma}} c_{\lambda' k' \bar{\sigma}}\rangle \langle\langle  d_{\sigma} |d^{\dagger}_{\sigma}\rangle\rangle _{E}. 
\end{equation}
The motivation for the decoupling comes from the knowledge that
the many-body Kondo singlet observed between the electron localized on the dot and the conduction 
electrons is due to anti-ferromagnetic spin flip processes. Thus performing this decoupling, we concentrate
on the spin, say up, particle moving in a self-consistently (see below) determined dynamic field of the  
spin down particles tunneling between the dot and electrodes.
These  approximations are known as Lacroix decoupling \cite{lacroix1981}. They are valid up to the 
second order in the coupling $V_{\lambda k \sigma}$ to the leads. The above approximations result
in the appearance of the various self-energies given as follows \cite{lavagna2015}:
\begin{eqnarray}
b_{1\bar{\sigma}}(E) &=& \sum_{\lambda k}\frac{V^{*}_{\lambda k \bar{\sigma}}
\langle d^{\dagger}_{\bar{\sigma}} c_{\lambda k \bar{\sigma}}\rangle}{E-\varepsilon_{\lambda k}-\varepsilon_1 +i\tilde{\gamma}^{\bar{\sigma}}_1}, \\
b_{2\bar{\sigma}}(E) &=& \sum_{\lambda k}\frac{V^{*}_{\lambda k \bar{\sigma}}
\langle c^{\dagger}_{\lambda k \bar{\sigma}}d_{\bar{\sigma}} \rangle}{E-\varepsilon_{\lambda k}-\varepsilon_1 +i\tilde{\gamma}_2} \\
\Sigma^{(1)}_{\bar{\sigma}}(E) &=& \sum_{\lambda k}\frac{|V_{\lambda k\bar{\sigma}}|^{2}}{E-\varepsilon_{\lambda k}-\varepsilon_{1}+i\tilde{\gamma}^{\bar{\sigma}}_1}\\
\Sigma^{(2)}_{\bar{\sigma}}(E) &=&\sum_{\lambda k}\frac{|V_{\lambda k\bar{\sigma}}|^{2}}{E+\varepsilon_{\lambda k}-\varepsilon_{2}+i\tilde{\gamma}_2}
\end{eqnarray}
and 
\begin{eqnarray}
\Sigma^{T}_{1\bar{\sigma}}(E)
&=& \sum_{\lambda k} \sum_{\lambda' k'} \frac{V^{*}_{\lambda k \bar{\sigma}} V_{\lambda' k' \bar{\sigma}} 
\langle c^{\dagger}_{\lambda' k' \bar{\sigma} } c_{\lambda k \bar{\sigma}}\rangle}{E-\varepsilon_{\lambda k} -\varepsilon_1 +i\tilde{\gamma}^{\bar{\sigma}}_1}, \\
\Sigma^{T}_{2\bar{\sigma}}(E) &=& \sum_{\lambda k}\sum_{\lambda' k'} \frac{V_{\lambda k\bar{\sigma}} V^{*}_{\lambda' k'\bar{\sigma}}
 \langle c^{\dagger}_{\lambda k\bar{\sigma}} c_{\lambda' k' \bar{\sigma}}\rangle}
{E+\varepsilon_{\lambda k}-\varepsilon_2 +i\tilde{\gamma}_2}.
\end{eqnarray} 
They can easily be expressed in terms of the retarded ($r$) and advanced ($a$) Green functions 
$\langle\langle d_\sigma|d^\dagger_\sigma\rangle\rangle^{r,a}_E$ of the opposite spin
GFs. In these expressions we have introduced two important 
generalisations as proposed earlier \cite{lavagna2015}. They are suggested by the work
of Van Roermund et al.\ who have extended the EOM technique and systematically studied the Anderson model up to the
fourth order \cite{vanroermund2010}, $i.e.$, up to $|V_{\lambda \vec{k}\sigma}|^4$. One the main findings of that paper is
that the primary effect of higher order processes is to provide a finite lifetime of excited states on the dot, 
and a renormalization of the on-dot energy level $\varepsilon_d$. These two effects are taken into account by
replacing the imaginary part in the GFs, $i0$, by $i\tilde{\gamma}_\alpha$, and $\varepsilon_d$ 
by $\tilde{\varepsilon}_d$. Finally we obtain the following expressions:
\begin{equation}
b_{1\bar{\sigma}}(E) = \int \frac{d\varepsilon}{2\pi}\frac {\sum_{\lambda} \Gamma^{\lambda}_{\bar{\sigma}} f_{\lambda} (\varepsilon) 
\langle\langle d_{\bar{\sigma}}|d^{\dagger}_{\bar{\sigma}}\rangle\rangle ^{a}_{\varepsilon}}
{E-\varepsilon-\varepsilon_1+i\tilde{\gamma}^{\bar{\sigma}}_1},
\end{equation}
\begin{equation}
b_{2\bar{\sigma}}(E) = \int \frac{d\varepsilon}{2\pi}\frac {\sum_{\lambda} \Gamma^{\lambda}_{\bar{\sigma}} f_{\lambda} (\varepsilon) 
\langle\langle d_{\bar{\sigma}}|d^{\dagger}_{\bar{\sigma}}\rangle\rangle ^{a}_{\varepsilon}}
{E+\varepsilon-\varepsilon_2+i\tilde{\gamma}^{\bar{\sigma}}_2},
\end{equation}
\begin{equation}
\Sigma^{T}_{1\sigma}(E)=\int\frac{d\varepsilon}{2\pi}\frac{\sum_{\lambda}\Gamma^{\lambda}_{\bar{\sigma}} f_{\lambda}(\varepsilon)[1+\frac{i}{2}
\Gamma_{\bar{\sigma}}\langle\langle d_{\bar{\sigma}}|d^{\dagger}_{\bar{\sigma}}\rangle\rangle^{a}_{\varepsilon}]}
{E-\varepsilon-\varepsilon_1+i\tilde{\gamma}^{\bar{\sigma}}_1},
\end{equation}
\begin{equation}
\Sigma^{T}_{2\bar{\sigma}}(E) =\int \frac{d\varepsilon}{2\pi} \frac{\sum_\lambda \Gamma^{\lambda}_{\bar{\sigma}} 
 f_{\lambda}(\varepsilon) [1-\frac{i}{2}\Gamma_{\bar{\sigma}}\langle\langle d_{\bar{\sigma}}|d^{\dagger}_{\bar{\sigma}}\rangle\rangle^{r}_{\varepsilon}]}{E +\varepsilon - \varepsilon_2+ i\tilde{\gamma}_2}.
\end{equation}
In the above we have introduced $\varepsilon_1=\tilde{\varepsilon}_\sigma-\tilde{\varepsilon}_{\bar{\sigma}}$,
and $\varepsilon_2=\tilde{\varepsilon}_\sigma+\tilde{\varepsilon}_{\bar{\sigma}}+U$. The subscripts 1 and 2 refer to
the excited 1- and 2-electron states of the dot, respectively. The self-energies are 
calculated iteratively and self-consistently with the GF (\ref{sol-gf}). 
The remaining two self-energies $\Sigma_\sigma^{(1,2)}$ are equal to $\Sigma_{0\sigma}$ 
for $i\tilde{\gamma}^\alpha_{1,2}=i0^+$; however, for arbitrary values of $i\tilde{\gamma}^\alpha_{1,2}$ they have 
to be calculated directly from the definition,
\begin{equation}
\Sigma^{(1,2)}_\sigma(E)=\sum_{\lambda} \Gamma^{\lambda}_{\bar{\sigma}}\int \frac{d\varepsilon}{2\pi}\frac {1}
{E \mp \varepsilon-\varepsilon_{1,2}+i\tilde{\gamma}^{\bar{\sigma}}_{1,2}}.
\end{equation}

\end{document}